	\documentclass[%
	reprint,
	superscriptaddress,
	showpacs,preprintnumbers,
	amsmath,amssymb,
	aps,
	prx,
	longbibliography
	]{revtex4-2}
	\usepackage{color}
	\usepackage{graphicx}
	\usepackage{bm}
	\usepackage{mathrsfs}
	\usepackage{dsfont}
    \usepackage[unicode=true, breaklinks=false, pdfborder={0 0 1}, backref=false, colorlinks=true, linkcolor=blue, citecolor=blue]{hyperref}
	
	\makeatletter
	
	\newcommand{\Rmnum}[1]{\expandafter\@slowromancap\romannumeral #1@}
	
	\makeatother

\begin{document}
	\title{Security analysis method for practical quantum key distribution with arbitrary encoding schemes}
	
	\author{Zehong Chang}
	\affiliation{National Local Joint Engineering Research Center for Precision Surgery \& Regenerative Medicine, Shaanxi Provincial Center for Regenerative Medicine and Surgical Engineering, The First Affiliated Hospital of Xi’an Jiaotong University, Xi’an, Shaanxi Province 710061, China}
	\affiliation{Ministry of Education Key Laboratory for Nonequilibrium Synthesis and Modulation of Condensed Matter, Shaanxi Province Key Laboratory of Quantum Information and Quantum Optoelectronic Devices, School of Physics, Xi’an Jiaotong University, Xi’an 710049, China}

	\author{Fumin Wang}
	\email{wh2009616@163.com}
	\affiliation{National Local Joint Engineering Research Center for Precision Surgery \& Regenerative Medicine, Shaanxi Provincial Center for Regenerative Medicine and Surgical Engineering, The First Affiliated Hospital of Xi’an Jiaotong University, Xi’an, Shaanxi Province 710061, China}
	\affiliation{Ministry of Education Key Laboratory for Nonequilibrium Synthesis and Modulation of Condensed Matter, Shaanxi Province Key Laboratory of Quantum Information and Quantum Optoelectronic Devices, School of Physics, Xi’an Jiaotong University, Xi’an 710049, China}

	\author{Xiaoli Wang}
	\affiliation{Ministry of Education Key Laboratory for Nonequilibrium Synthesis and Modulation of Condensed Matter, Shaanxi Province Key Laboratory of Quantum Information and Quantum Optoelectronic Devices, School of Physics, Xi’an Jiaotong University, Xi’an 710049, China}
	
	\author{Xiaofei Liu}
	\affiliation{National Local Joint Engineering Research Center for Precision Surgery \& Regenerative Medicine, Shaanxi Provincial Center for Regenerative Medicine and Surgical Engineering, The First Affiliated Hospital of Xi’an Jiaotong University, Xi’an, Shaanxi Province 710061, China}
	
	\author{Rongqian Wu}
	\affiliation{National Local Joint Engineering Research Center for Precision Surgery \& Regenerative Medicine, Shaanxi Provincial Center for Regenerative Medicine and Surgical Engineering, The First Affiliated Hospital of Xi’an Jiaotong University, Xi’an, Shaanxi Province 710061, China}
	
	\author{Yi lv}
	\affiliation{National Local Joint Engineering Research Center for Precision Surgery \& Regenerative Medicine, Shaanxi Provincial Center for Regenerative Medicine and Surgical Engineering, The First Affiliated Hospital of Xi’an Jiaotong University, Xi’an, Shaanxi Province 710061, China}
		
	\author{Pei Zhang}
	\email{zhangpei@mail.ustc.edu.cn}
	\affiliation{Ministry of Education Key Laboratory for Nonequilibrium Synthesis and Modulation of Condensed Matter, Shaanxi Province Key Laboratory of Quantum Information and Quantum Optoelectronic Devices, School of Physics, Xi’an Jiaotong University, Xi’an 710049, China}
	\date{\today}
	
	\begin{abstract}
	Quantum key distribution (QKD) gradually has become a crucial element of practical secure communication. In different scenarios, the security analysis of genuine QKD systems is complicated. A universal secret key rate calculation method, used for realistic factors such as multiple degrees of freedom encoding, asymmetric protocol structures, equipment flaws, environmental noise, and so on, is still lacking. Based on the correlations of statistical data, we propose a security analysis method without restriction on encoding schemes. This method makes a trade-off between applicability and accuracy, which can effectively analyze various existing QKD systems. We illustrate its ability by analyzing source flaws and a high-dimensional asymmetric protocol. Results imply that our method can give tighter bounds than the Gottesman-Lo-L{\"u}tkenhaus-Preskill (GLLP) analysis and is beneficial to analyze protocols with complex encoding structures. Our work has the potential to become a reference standard for the security analysis of practical QKD.
	\end{abstract}
	
	\maketitle  
	\section{Introduction} 
	\label{sec:introduction}
	The security of classical cryptographic schemes, based on the restrictions of computing resources and strategies, is threatened by quantum computers and algorithms. Guaranteed by the principles of quantum mechanics, quantum key distribution (QKD) can provide unconditional security and against attacks from quantum computers \cite{osti_5184714,PhysRevLett.67.661,Lo2050,PhysRevLett.85.441,PhysRevA.72.012332,Koashi_2006}. QKD has developed significantly over the past three decades \cite{Chen2021AnIS,PhysRevLett.126.250502,proietti2021experimental,zhong2021experiment}, and more efforts are required for the ultimate goal of a global QKD network.
	
	Compared with classical cryptography, the lower secure key rates of QKD pose a challenge to its widespread adoption over long distances. A promising approach is accessing high-dimensional Hilbert spaces \cite{PhysRevLett.88.127902,PhysRevLett.89.240401,Gr_blacher_2006}, such as encoding with the spatial modes \cite{PhysRevA.88.032305,Mirhosseini_2015}. This method can increase channel capacity and resistance to noise. Another novel approach is using the twin-filed (TF) QKD, that breaks the fundamental linear rate-distance limit through single-photon interference \cite{TF-QKD,PhysRevX.8.031043,PhysRevA.98.062323}. These developments of theory promotes the construction of QKD networks. But there are challenges in the practical implementation \cite{RevModPhys.74.145,RevModPhys.81.1301,RevModPhys.92.025002}. On the one hand, due to imperfect equipment \cite{1365172}, there exist deviations between implementation and theory, such as flawed and leaky sources. On the other hand, environmental factors also limits the performance of protocol, such as the misalignment of reference frame \cite{PhysRevLett.107.110501,dong2014attack} and turbulance \cite{PhysRevLett.94.153901}. Analyzing above issues is significant to inspire the design of practical protocols as well as avoid potential security loopholes.
	
	Some works have made efforts to analyze the security of QKD with imperfect equipment. The standard Gottesman-Lo-L{\"u}tkenhaus-Preskill (GLLP) security proof \cite{1365172} allows one to address these problems conservatively. This analysis leads to a low achievable secret key rate and is not robust against channel losses. In 2014, Tamaki et al. \cite{tamaki2014loss} proposed a loss-tolerant protocol by using the basis mismatch events to precisely bound the phase error rate. This protocol requires full characterization of imperfect qubit states \cite{PhysRevA.90.052319} and is difficult to analyze complex encoding schemes. 
	
	In 2016, Coles et al. proposed a numerical method to analyze the unstructured protocols \cite{Coles}, including encoding with qudit, non-orthogonal states \cite{PhysRevLett.68.3121} and non-mutually unbiased bases (non-MUBs) \cite{Matsumoto_2010}. The method focuses more on the applicability of numerical methods and lacks a general discussion on other realistic factors, such as equipment flaws and environmental noises. Besides, Sun et al. \cite{PhysRevApplied.12.034039} analyzed the reference frame independent (RFI) protocol \cite{PhysRevA.82.012304} with fewer states, which is a typical case to analyze the encoding flaws and misalignment simultaneously. This protocol is two-dimensional and its protocol structure is limited. Furthermore, some recent works using numerical methods have focused on solving practical problems such as the finite-key analysis and mismatched detection \cite{numericalnpj,PhysRevResearch.3.013274}. The potential of numerical methods has not yet been fully exploited.
	
	In this work, we provide a security analysis method for arbitrary encoding schemes considering practical issues. We demonstrate its accuracy, compared with GLLP analysis, in two examples with different source flaws. In the first example of the state-dependent flaw, our method shows better tolerance for modulation errors. Then, in the example of TF-QKD with non-random phases, our method still outperforms GLLP analysis without considering flaws enhancement caused by channel losses. Furthermore, we apply this method to the high-dimensional asymmetric encoding schemes. In particular, we adopt the mutually partially unbiased bases (MPUB) protocol \cite{PhysRevA.101.032340} which encodes with multiple degrees of freedom of spatial modes. We propose a  decoy-state RFI-MPUB protocol and its results exhibit a comparable performance compared to the ideal $d=4$ BB84 protocol. Our analyses imply that this method can comprehensively analyze various complex factors in practical QKD.
	
	\section{Theory and Models for Security analysis} 
	\label{sec: main results}
\begin{figure}[!t]
		\centering
		\includegraphics[width=8.6cm]{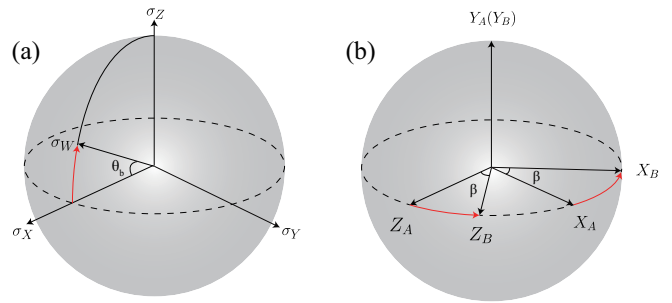}
		\caption{The models of basis bias and the misalignment of reference frame. The basis bias angle $\theta_{b}$ exists in $X$-$Z$ plane shown in (a), where $W$ basis is rotated away from the ideal $X$ basis. The three Pauli operators are $\{\sigma_{Z},\sigma_{X},\sigma_{Y}\}$. Considering the misalignment of the reference frame in (b), the coordinate axes on Alice's side and Bob's side are expressed with subscripts A and B, denote as $\{Z_{A}, W_{A}, Y_{A}\}$ and $\{Z_{B}, W_{B}, Y_{B}\}$ respectively. The $Y$ basis is well-defined in (b) with the rotation angle $\beta$. The directions of rotations are indicated by red arrows. Here we distinguish between the Bloch sphere of quantum state space and the reference frame space.}
		\label{fig:C1}
	\end{figure}

	Affected by application scenarios and practical issues, the encoding schemes are diverse, mainly reflected in dimension and symmetry (complementarity between bases). Here, the asymmetric structure and misalignment are modeled as different parameters. We define the deviation from quantum states of MUBs as bias angles. In a $d$-dimensional QKD protocol, its maximum bias angle $\theta\in \left[0,\pi/4 \right]$ between the states of two bases $\{k,l\}$ is:
	\begin{align}
		\label{main1}
		\theta=\frac{1}{2}\arcsin(\max\{\left|\langle\varphi^{m}_{k}\mid\varphi^{n}_{l}\rangle\right|^{2}-\mid\langle\varphi^{m}_{k}\mid\varphi^{p\neq n}_{l}\rangle\mid^{2} \}) ,
	\end{align}
	\begin{align}
		\label{main2}
		\theta=\frac{1}{2}(\max\{\left|ln \langle\varphi^{m}_{k}\mid\varphi^{n}_{l}\rangle\right|-\mid ln \langle\varphi^{m}_{k}\mid\varphi^{p\neq n}_{l}\rangle\mid \}) ,
	\end{align}
	where $|\varphi^{\kappa}_{j}\rangle$ represents the $\kappa$-th state in basis $j$ and $\kappa\in[0,d-1]$. This definition is dimension-independent and gives the maximum deviation estimate. Eq. (\ref{main1}) and Eq. (\ref{main2}) represent encoding with a single-photon source and a weak coherent source respectively. We assume this bias lies on the $X$-$Z$ plane of Bloch sphere in two-dimensional encoding protocol, if not, it can be transformed using a filter operation \cite{tamaki2014loss}. State bias is reflected in security loopholes caused by lacking of complementarity between bases in the phase error estimation \cite{koashi2007complementarity}.
	Hence, basis bias angle, defined as $\theta_{b}=2\theta$, is used in the following work as shown in Fig. \ref{fig:C1}(a). We use the reference frame rotation model of RFI protocol in Fig. \ref{fig:C1}(b), suitable for chip-to-chip \cite{PhysRevLett.112.130501} and earth-to-satellite QKD.
	
	According to these models, we modify the error rates of quantum channels. A generalized form of our theory:
	\begin{align}
		\label{main3}
		e_{ij} &=\dfrac{1-M(1-2e_{exp})}{2},
	\end{align}
	\begin{align}
		\label{main4}
		M=\dfrac{1}{\sqrt{1+\sin^{2}\theta_{b}}},
	\end{align}
	where $e_{ij}$ denotes the modified error rates which are defined as a fictitious bit-error rate when Alice and Bob measure in $i$ basis and $j$ basis respectively. $e_{exp}=(1-N_{ij}(1-2Q))/2$ is the error rate measured experimentally. $N_{ij}$ is related to the probability of an event occurring, used for theoretical calculations. $Q$ is quantum bit error rate (QBER) in an ideal case. Derived from the correlations between bases, $M$ maintains completeness of systems. 
	
	Generally, a protocol can be determined by the key generation and the experimental constraints used for parameter estimation. For simplicity, we assume only one basis is used for the key generation with measurements  $\{Z_{A}^{j}\}$. And the modified error rates are used to constraint Eve's available information. Next, we will use this theory to quickly analyze various issues in actual QKD based on the numerical optimization method in Ref. \cite{Coles} (Details in supplemental material).
	
	\section{BASIS-DEPENDENT FLAW}
	One main issue of sources in a practical QKD system is the basis-dependent flaw that stems from the discrepancy of density matrices corresponding to two bases. In phase encoding schemes, this discrepancy comes from the modulation errors of phase modulator. The difference between actual phases and expected cases is defined as $\delta$ which can be experimentally measured or tested \cite{xu2015experimental}.
	
	Here, we employ the vacuum+weak decoy-state method \cite{ma2005practical} to illustrate our analysis. We select two typical modulation error values: $\delta\leq0.062$ corresponds to the modulation error with an advanced phase-stabilized interferometer with about $99.9\%$ visibility; $\delta=0.127$ corresponds to the upper bound of modulation errors of commercial plug$\&$play systems with its fidelity $99.81\%$. The key generation rate is given by:
	\begin{align}
	\label{flaw1}
	R\geq q \{-f(E_{\mu})Q_{\mu}h(E_{\mu})+Q_{1}(1-h(e^{U}_{1}))\},
	\end{align}
	where $q$ depends on the implementation ($q\approx 1$ for an efficient BB84 protocol \cite{2005Lo}). $h(x)=-x\log_{2}(x/(d-1))-(1-x)\log_{2}(1-x)$ is the binary entropy function in dimension $d$. $\mu$ is the intensity of signal states. The gain $Q_{\mu}$ and QBER $E_{\mu}$ of signal states can be measured in experiments. $Q_{1}=Y_{1}^{L}\mu e^{-\mu}$, $Y_{1}^{L}$ and $e_{1}^{U}$ can be estimated with the approach in Ref. \cite{ma2005practical}. And the mutual information $H(Z_{A}|E)$, the result of numerical optimization, corresponds to $1-h(e^{U}_{1})$.
	
	Based on our analysis, measurements used to generate the raw key are $Z_{A}^{j}=\{|0\rangle\langle0|,|1\rangle\langle1|\}$ and the constraints used for optimization problem are as follows:
	\begin{align}
	\label{flaw2}
	\langle I \rangle =1,
	\end{align}
	\begin{align}
	\label{flaw3}
	\langle E_{Z} \rangle =(1-M(1-2e_{1}^{U}))/2,
	\end{align}
	\begin{align}
	\label{flaw4}
	\langle E_{X,mod} \rangle =(1-M(1-2e_{1}^{U}))/2,
	\end{align}
	where operators $E_{Z}:=(I-\sigma^{A}_{z}\otimes\sigma^{B}_{z})/2$, $E_{X,mod}:=(I-\sigma^{A}_{w}\otimes\sigma^{B}_{w})/2$ and $\sigma^{A(B)}_{w}=\cos\theta_{b}\sigma_{X}+\sin\theta_{b}\sigma_{Z}$. Moreover, the standard GLLP security analysis for BB84 with source flaws, its phase error with the correction is:
	\begin{align}
	\label{flaw5}
	e^{U}_{phase} &=e^{U}_{1}+4\Delta^{\prime}(1-\Delta^{\prime})(1-2e^{U}_{1}) \nonumber \\
	&+4(1-2\Delta^{\prime})\sqrt{\Delta^{\prime}(1-\Delta^{\prime})e^{U}_{1}(1-e^{U}_{1})} \nonumber \\
	&\leq e^{U}_{1}+4\Delta^{\prime} +4\sqrt{\Delta^{\prime} e^{U}_{1}},
	\end{align}
	where $\Delta^{\prime}\leq \Delta / Y_{1}^{L}$ is the balance of a quantum coin. This bias is enhanced during the channel losses. The imperfect of fidelity between density matrices is $\Delta=(1-F(\rho_{z}, \rho_{x}))/2$. 
	
\begin{figure}
		\centering
		\includegraphics[width=8.6cm]{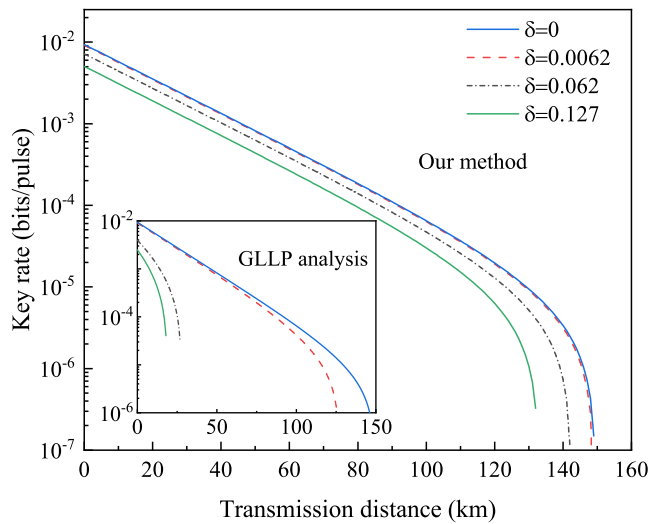}
		\caption{ Practical key rates with source flaws. $\delta$ is the upper bound of phase modulation errors. Parameters used for numerical simulations: $\mu=0.48$, $\nu=0.13$, $\omega=0$; background count rate, $Y_{0} = 1.7\times10^{-6}$; loss coefficient of the channel, $\alpha= 0.21$ $km/dB$; intrinsic detector error probability, $e_{d}=3.3\%$; detection efficiency, $\eta_{B}=4.5\%$; the efficiency of error correction $f(E_{\mu})=1.22$. The bias angle is $\theta_{b}=2\delta$.}
		\label{fig:C2}
	\end{figure}

	Our simulation results display in Fig. \ref{fig:C2}. The GLLP analysis pessimistically assumes Eve can enhance the flaw by exploiting channel losses, as shown in the inset figure. As a comparison, our analysis can substantially outperform GLLP in different cases. The security analysis based on our theory with a commercial system $\delta=0.127$ (green solid curve) can be made secure over $120$ km, while the maximal distance with GLLP is $20$ km. However, our theory is not tightest compared with the loss-tolerant protocol. In Ref. \cite{xu2015experimental}, the key rate of loss-tolerant BB84 protocol with $\delta=0.134$ can be almost the same as the case $\delta=0$. The main reason is that our constraints are coarse-grained, hence, our key rates can be higher if we accept basis mismatched data for constraints, as discussed in Ref. \cite{Coles}. Besides, we estimate the mutual information between Alice and Eve by maximizing the impact of source flaws in channels. This is a small but inevitable error.
	
	\section{NON-RANDOM PHASES} 
	\label{sec:example 2}
	Another important imperfection of sources in a practical QKD system is the non-random phases \cite{lo2006security}. The eavesdropper has a priori knowledge about phases of signal states, such as the original TF-QKD. Users announce the phase information in communication and employ a coherent encoding scheme rather than the decoy-state method in this protocol. Its information-theoretic security was firstly proved in Ref. \cite{tamaki2018information} by TF$^{*}$-QKD which using the decoy-state method in Test mode and the weak coherent encoding in Code mode. This proof takes its imbalance of a quantum coin into account and modifies the phase error. A difference of GLLP analysis here is that the enhancement caused by post-selection is not dependent on the channel losses.
	
	Here, we analyze this flaw based on this TF$^{*}$-QKD. Considering the infinite decoy states method \cite{PhysRevLett.91.057901} in Test mode, we can write its asymptotic key rate as:
		 \begin{align}
	 	\label{random1}
	 	R=q\{-f(E_{\mu})Q_{\mu}h(E_{\mu})+Q_{1}(1-h(e^{U}_{1}))\}.
	\end{align}
	Here $q=1/2$ and the modified phase error rate of GLLP analysis can calculate from Eq. (\ref{flaw5}) with $\Delta=2\pi/8$. According to the post-selection in TF$^{*}$-QKD and Eq. (\ref{main2}), we also assume the maximum bias angle is $\theta_{b}=2\pi/8$ and constraints are the same as Eq. (\ref{flaw2})-(\ref{flaw4}).
	\begin{figure}
		\centering
		\includegraphics[width=8.6cm]{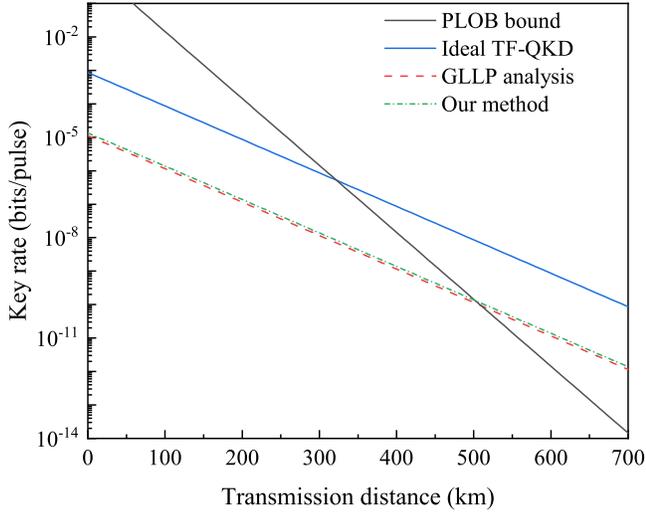}
		\caption{Key rates of TF$^{*}$-QKD protocol with the non-random phases. Parameters used for numerical simulations: $\mu_{A}=\mu_{B}=0.0012$; dark counts rate, $P_{d} = 10^{-11}$; loss coefficient of the channel, $\alpha= 0.2$ $km/dB$; detection efficiency, $\eta=80\%$; phase slices, $M_{1}=16$; the efficiency of error correction $f(E_{\mu})=1.1$.} 
		\label{fig:C3}
	\end{figure}
	
	We plot the resulting key rates in Fig. \ref{fig:C3}. TF$^{*}$-QKD is almost optimal at 500 km. As a comparison, the key rates with GLLP analysis (red dash line) are slightly lower than cases based on our method (green short dash dot line). The curves are close because the GLLP analysis here does not depend on channel losses. Both of analyses clearly show the $\sqrt{\eta}$ scaling, which makes it possible to overcome the secret key capacity limit (black solid line). 
	
	\section{HIGH-DIMENSIONAL asymmetric ENCODING} 
	\label{sec:example 1}
	The above analyses are two-dimensional protocols with shared reference frames. Their misalignment comes from the alignment and stability of optical systems \cite{ma2005practical}. It is considered as an inherent and known error in the decoy-state method. Hence, we analyze the high-dimensional spatial modes encoding scheme in free-space link with unknown misalignment. The MPUB protocol extremely simplifies the implementation \cite{BEIJERSBERGEN1993123,Jia:18} and enhances the robustness to turbulence by using non-MUBs encoded with multiple degrees of freedom. The security analysis in Ref. \cite{PhysRevA.101.032340} is not generalizable to practical issues. Here, we propose a decoy-state RFI-MPUB protocol to close the gaps between assumptions made in security proofs and actual implementations (Details in supplemental material).
	
	The dimension of this protocol is $d=N+1$ with its spatial mode order $N=n+m$. We adopt three bases structure \cite{Br_dler_2016} to encode information. The Laguerre-Gaussian ($LG$) basis is used for generating raw key rates. And the same mode order Hermite-Gaussian ($HG$) modes and $HG^{\urcorner}$ modes ($HG$ modes rotated by $45^{\circ}$) are used for parameter estimation. Because the RFI protocol does not limit the form of $\rho_{AB}$ strictly, the protocol with asymmetric structure is still applicable \cite{PhysRevA.92.042319}. Combined with our method, the statistical parameter still satisfy:
	\begin{align}
		\label{HD1}
		P &:=\sum_{i,j}  (1-2e_{ij})^{2}\leq2,
	\end{align}
	where $P$ is independent of the misalignment of reference frame when there are no eavesdroppers and other issues \cite{Sheridan_2010,PhysRevA.94.062330}. For simplicity, we calculate Eve’s information directly with $H(Z_{A}|E)=log_{2}d-I_{E}$. The key-map operators in $LG$ basis are $\{|l_{i}\rangle\langle l_{i}|, i\in[0,d-1]\}$. In the following, we will calculate the error operators. In RFI-MPUB protocol, arbitrary encoding states $|\alpha_{n,m}\rangle$ can be decomposed into a set of $HG$ modes states ${|h_{N-k,k}\rangle}$ \cite{BEIJERSBERGEN1993123}:
	\begin{align}
	   	\label{HD2}
	   	|\alpha_{n,m}\rangle = U_{q}|h_{N-k,k}\rangle,\quad
	   	U_{q}:=\sum_{k=0}^{N} (i^{k})^{q}b(n,m,k).
	\end{align}
	where the integer number $k\in[0,N]$. $U_{q}$ is a transformation matrix, and $\{q=0,1\}$ represent the  $HG^{\urcorner}$ modes and LG modes respectively. In addition, we define $U_{2}$ as an identity matrix. The real coefficients  $b(n,m,k)$ is:
	\begin{align}
	   	\label{HD3}
	   	b(n,m,k)=(\dfrac{(N-k)!k!}{2^{N}n!m!})^{1/2}\dfrac{1}{k!}\dfrac{d^{k}}{dt^{k}}[(1-t)^{n}(1+t)^{m}]_{t=0},
	\end{align}
	where the factor $i^{k}$ corresponds to a $\pi/2$ relative phase difference between successive components. The error operator is defined as $E_{q}:=\openone-C_{q}$, where $q=\{0,1,2\}$ represent for $\{HG^{\urcorner}, LG, HG\}$ respectively.
	\begin{align}
	   	\label{HD4}
	   	C_{q}=\sum_{{n,m}}U_{q}|h_{n,m}\rangle\langle h_{n,m}|U_{q}^{\dagger}\otimes U_{q}^{\dagger}|h_{n,m}\rangle\langle h_{n,m}|U_{q}.
	\end{align}
	 Finally, the constraints can be represented as:
	   	\begin{align}
	   	\label{HD5}
	   	\langle I\rangle = 1,
	\end{align}
	\begin{align}
	   	\label{HD6}
	   	\langle E_{key}\rangle=e^{U}_{1},
	\end{align}
	\begin{align}
	   	\label{HD7}
	   	\langle E_{est}\rangle=(1-M(1-2e^{U}_{1}))/2.
	\end{align}
	Here $E_{key}=E_{1}$ and $E_{est}=(E_{0}+E_{2})/2$. We use the same parameter settings in Ref. \cite{Wang:21}. And the key rates in the asymptotic case are given by Eq. (\ref{flaw1}).
	
\begin{figure}
		\centering
		\includegraphics[width=8.6cm]{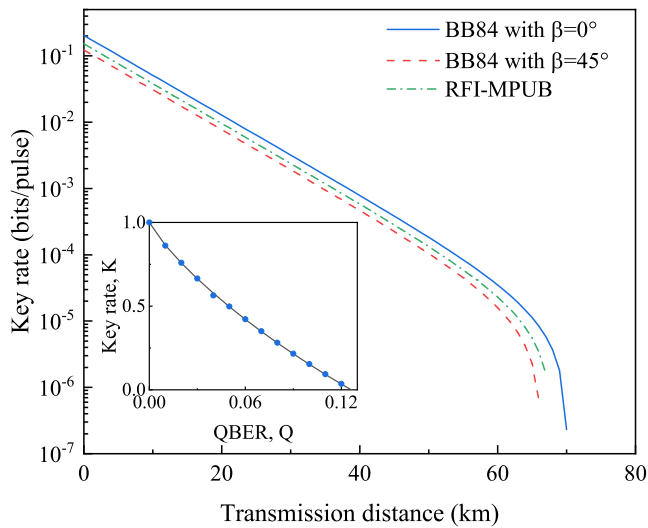}
		\caption{Key rates performance of RFI-MPUB and $d=4$ BB84 protocol. Parameters used for numerical simulations: $\mu=0.3$, $\nu=0.05$, $\omega=0$; background count rate, $Y_{0} = 3\times10^{-6}$; loss coefficient of the channel, $\alpha= 0.6$ $km/dB$; intrinsic detector error probability, $e_{d}=1.5\%$; detection efficiency, $\eta_{B}=50\%$; the efficiency of error correction $f(E_{\mu})=1.22$; basis bias angle calculated from Eq. (\ref{main1}) $\theta_{b} \approx14.48^{\circ}$ .}
		\label{fig:C4}
	\end{figure}

	In Fig. \ref{fig:C4}, the raw key rates and maximum transmission distance of RFI-MPUB (green dash dot curve) can be comparable with the ideal BB84 protocol (black solid curve). An overall decrease of key rates achieved in RFI-MPUB can be explained by the leak of information dues to basis bias in actual environment. However, the key rates are still higher than BB84 protocol with misalignment angle $\beta=45^{\circ}$ (red dash curve) which might be worse in actual environment. In the inset, output results of our approximation (blue dots) are consistent with the RFI protocol (black line), and the deviation of data comes from accuracy of program calculation. 
	
	\section{Conclusion and Discussion} 
	\label{sec:Conclusions}
	In conclusion, we propose a security analysis method by modeling asymmetric structures of protocols and misalignment of reference frames. We extract a factor from correlations of statistical results and apply it to analyze various issues in different scenarios. 
	For source flaws problems with the same amount of experimental data, our method can obtain higher key rates than GLLP analysis. Moreover, this method overcomes the restriction of encoding schemes, hence, is a powerful tool for security analysis of protocols with asymmetric structures or qudit encoding. 
	
	Due to the overestimation of bias angle, our method sacrifices some accuracy, but it achieves wider applicability compared with previous methods. In principle, our analysis is slightly lower than the loss-tolerant protocol with the same data. The ability of loss tolerance in our method can be enhanced by accepting mismatched data. We envision that our method can be a reference standard for QKD scientists. In future work, we hope to extend our method to other realistic factors or scenarios, such as the finite-key analysis \cite{PhysRevLett.126.100501,PhysRevResearch.3.023019} and high-dimensional conference key agreement protocols \cite{murta2020quantum}. Furthermore, our analysis may be instructive for other different types of asymmetric structure \cite{PhysRevA.97.042347,Lim2021}.
	
	\section*{Acknowledgments}
	Fumin Wang and Zehong Chang contributed equally to this manuscript. We thank Shihao Ru for linguistic assistance during the preparation of this manuscript.
	
	This work was in part supported by the National Nature Science Foundation of China(Grant Nos. 11804271, 91736104 and 12074307). In addition, Yi Lv acknowledges support from National Key R\&D Project of China (Nos. 2018YFC0115300 and 2018YFC0115305, YL);  National Natural Science Foundation of China (No.81727802) and Innovation Capacity Support Plan of Shaanxi Province (No. 2020TD-040, RW). 
	
\nocite{*}


\begin{thebibliography}{55}%
	\makeatletter
	\providecommand \@ifxundefined [1]{%
		\@ifx{#1\undefined}
	}%
	\providecommand \@ifnum [1]{%
		\ifnum #1\expandafter \@firstoftwo
		\else \expandafter \@secondoftwo
		\fi
	}%
	\providecommand \@ifx [1]{%
		\ifx #1\expandafter \@firstoftwo
		\else \expandafter \@secondoftwo
		\fi
	}%
	\providecommand \natexlab [1]{#1}%
	\providecommand \enquote  [1]{``#1''}%
	\providecommand \bibnamefont  [1]{#1}%
	\providecommand \bibfnamefont [1]{#1}%
	\providecommand \citenamefont [1]{#1}%
	\providecommand \href@noop [0]{\@secondoftwo}%
	\providecommand \href [0]{\begingroup \@sanitize@url \@href}%
	\providecommand \@href[1]{\@@startlink{#1}\@@href}%
	\providecommand \@@href[1]{\endgroup#1\@@endlink}%
	\providecommand \@sanitize@url [0]{\catcode `\\12\catcode `\$12\catcode
		`\&12\catcode `\#12\catcode `\^12\catcode `\_12\catcode `\%12\relax}%
	\providecommand \@@startlink[1]{}%
	\providecommand \@@endlink[0]{}%
	\providecommand \url  [0]{\begingroup\@sanitize@url \@url }%
	\providecommand \@url [1]{\endgroup\@href {#1}{\urlprefix }}%
	\providecommand \urlprefix  [0]{URL }%
	\providecommand \Eprint [0]{\href }%
	\providecommand \doibase [0]{https://doi.org/}%
	\providecommand \selectlanguage [0]{\@gobble}%
	\providecommand \bibinfo  [0]{\@secondoftwo}%
	\providecommand \bibfield  [0]{\@secondoftwo}%
	\providecommand \translation [1]{[#1]}%
	\providecommand \BibitemOpen [0]{}%
	\providecommand \bibitemStop [0]{}%
	\providecommand \bibitemNoStop [0]{.\EOS\space}%
	\providecommand \EOS [0]{\spacefactor3000\relax}%
	\providecommand \BibitemShut  [1]{\csname bibitem#1\endcsname}%
	\let\auto@bib@innerbib\@empty
	\bibitem [{\citenamefont {Bennett}\ and\ \citenamefont
		{Brassard}(1985)}]{osti_5184714}%
	\BibitemOpen
	\bibfield  {author} {\bibinfo {author} {\bibfnamefont {C.~H.}\ \bibnamefont
			{Bennett}}\ and\ \bibinfo {author} {\bibfnamefont {G.}~\bibnamefont
			{Brassard}},\ }\href {https://www.osti.gov/biblio/5184714} {\emph {\bibinfo
			{title} {Proceedings of the IEEE international conference on computers}}}\
	(\bibinfo  {publisher} {IEEE New York},\ \bibinfo {year} {1985})\BibitemShut
	{NoStop}%
	\bibitem [{\citenamefont {Ekert}(1991)}]{PhysRevLett.67.661}%
	\BibitemOpen
	\bibfield  {author} {\bibinfo {author} {\bibfnamefont {A.~K.}\ \bibnamefont
			{Ekert}},\ }\bibfield  {title} {\bibinfo {title} {Quantum cryptography based
			on bell's theorem},\ }\href {https://doi.org/10.1103/PhysRevLett.67.661}
	{\bibfield  {journal} {\bibinfo  {journal} {Phys. Rev. Lett.}\ }\textbf
		{\bibinfo {volume} {67}},\ \bibinfo {pages} {661} (\bibinfo {year}
		{1991})}\BibitemShut {NoStop}%
	\bibitem [{\citenamefont {Lo}\ and\ \citenamefont {Chau}(1999)}]{Lo2050}%
	\BibitemOpen
	\bibfield  {author} {\bibinfo {author} {\bibfnamefont {H.-K.}\ \bibnamefont
			{Lo}}\ and\ \bibinfo {author} {\bibfnamefont {H.~F.}\ \bibnamefont {Chau}},\
	}\bibfield  {title} {\bibinfo {title} {Unconditional security of quantum key
			distribution over arbitrarily long distances},\ }\href
	{https://doi.org/10.1126/science.283.5410.2050} {\bibfield  {journal}
		{\bibinfo  {journal} {Science}\ }\textbf {\bibinfo {volume} {283}},\ \bibinfo
		{pages} {2050} (\bibinfo {year} {1999})}\BibitemShut {NoStop}%
	\bibitem [{\citenamefont {Shor}\ and\ \citenamefont
		{Preskill}(2000)}]{PhysRevLett.85.441}%
	\BibitemOpen
	\bibfield  {author} {\bibinfo {author} {\bibfnamefont {P.~W.}\ \bibnamefont
			{Shor}}\ and\ \bibinfo {author} {\bibfnamefont {J.}~\bibnamefont
			{Preskill}},\ }\bibfield  {title} {\bibinfo {title} {Simple proof of security
			of the {BB84} quantum key distribution protocol},\ }\href
	{https://doi.org/10.1103/PhysRevLett.85.441} {\bibfield  {journal} {\bibinfo
			{journal} {Phys. Rev. Lett.}\ }\textbf {\bibinfo {volume} {85}},\ \bibinfo
		{pages} {441} (\bibinfo {year} {2000})}\BibitemShut {NoStop}%
	\bibitem [{\citenamefont {Renner}\ \emph {et~al.}(2005)\citenamefont {Renner},
		\citenamefont {Gisin},\ and\ \citenamefont {Kraus}}]{PhysRevA.72.012332}%
	\BibitemOpen
	\bibfield  {author} {\bibinfo {author} {\bibfnamefont {R.}~\bibnamefont
			{Renner}}, \bibinfo {author} {\bibfnamefont {N.}~\bibnamefont {Gisin}},\ and\
		\bibinfo {author} {\bibfnamefont {B.}~\bibnamefont {Kraus}},\ }\bibfield
	{title} {\bibinfo {title} {Information-theoretic security proof for
			quantum-key-distribution protocols},\ }\href
	{https://doi.org/10.1103/PhysRevA.72.012332} {\bibfield  {journal} {\bibinfo
			{journal} {Phys. Rev. A}\ }\textbf {\bibinfo {volume} {72}},\ \bibinfo
		{pages} {012332} (\bibinfo {year} {2005})}\BibitemShut {NoStop}%
	\bibitem [{\citenamefont {Koashi}(2006)}]{Koashi_2006}%
	\BibitemOpen
	\bibfield  {author} {\bibinfo {author} {\bibfnamefont {M.}~\bibnamefont
			{Koashi}},\ }\bibfield  {title} {\bibinfo {title} {Unconditional security of
			quantum key distribution and the uncertainty principle},\ }\href
	{https://doi.org/10.1088/1742-6596/36/1/016} {\bibfield  {journal} {\bibinfo
			{journal} {J Phys Conf Ser}\ }\textbf {\bibinfo {volume} {36}},\ \bibinfo
		{pages} {98} (\bibinfo {year} {2006})}\BibitemShut {NoStop}%
	\bibitem [{\citenamefont {Chen}\ \emph {et~al.}(2021)\citenamefont {Chen},
		\citenamefont {Zhang}, \citenamefont {Chen}, \citenamefont {Cai},
		\citenamefont {Liao}, \citenamefont {Zhang}, \citenamefont {Chen},
		\citenamefont {Yin}, \citenamefont {Ren}, \citenamefont {Chen}, \citenamefont
		{Han}, \citenamefont {Yu}, \citenamefont {Liang}, \citenamefont {Zhou},
		\citenamefont {Yuan}, \citenamefont {Zhao}, \citenamefont {Wang},
		\citenamefont {Jiang}, \citenamefont {Zhang}, \citenamefont {Liu},
		\citenamefont {Li}, \citenamefont {Shen}, \citenamefont {Cao}, \citenamefont
		{Lu}, \citenamefont {Shu}, \citenamefont {Wang}, \citenamefont {Li},
		\citenamefont {Liu}, \citenamefont {Xu}, \citenamefont {Wang}, \citenamefont
		{Peng},\ and\ \citenamefont {Pan}}]{Chen2021AnIS}%
	\BibitemOpen
	\bibfield  {author} {\bibinfo {author} {\bibfnamefont {Y.}~\bibnamefont
			{Chen}}, \bibinfo {author} {\bibfnamefont {Q.}~\bibnamefont {Zhang}},
		\bibinfo {author} {\bibfnamefont {T.-Y.}\ \bibnamefont {Chen}}, \bibinfo
		{author} {\bibfnamefont {W.}~\bibnamefont {Cai}}, \bibinfo {author}
		{\bibfnamefont {S.}~\bibnamefont {Liao}}, \bibinfo {author} {\bibfnamefont
			{J.}~\bibnamefont {Zhang}}, \bibinfo {author} {\bibfnamefont
			{K.}~\bibnamefont {Chen}}, \bibinfo {author} {\bibfnamefont {J.}~\bibnamefont
			{Yin}}, \bibinfo {author} {\bibfnamefont {J.-G.}\ \bibnamefont {Ren}},
		\bibinfo {author} {\bibfnamefont {Z.}~\bibnamefont {Chen}}, \bibinfo {author}
		{\bibfnamefont {S.-L.}\ \bibnamefont {Han}}, \bibinfo {author} {\bibfnamefont
			{Q.}~\bibnamefont {Yu}}, \bibinfo {author} {\bibfnamefont {K.}~\bibnamefont
			{Liang}}, \bibinfo {author} {\bibfnamefont {F.}~\bibnamefont {Zhou}},
		\bibinfo {author} {\bibfnamefont {X.}~\bibnamefont {Yuan}}, \bibinfo {author}
		{\bibfnamefont {M.}~\bibnamefont {Zhao}}, \bibinfo {author} {\bibfnamefont
			{T.-Y.}\ \bibnamefont {Wang}}, \bibinfo {author} {\bibfnamefont
			{X.}~\bibnamefont {Jiang}}, \bibinfo {author} {\bibfnamefont
			{L.}~\bibnamefont {Zhang}}, \bibinfo {author} {\bibfnamefont
			{W.}~\bibnamefont {Liu}}, \bibinfo {author} {\bibfnamefont {Y.}~\bibnamefont
			{Li}}, \bibinfo {author} {\bibfnamefont {Q.}~\bibnamefont {Shen}}, \bibinfo
		{author} {\bibfnamefont {Y.}~\bibnamefont {Cao}}, \bibinfo {author}
		{\bibfnamefont {C.-Y.}\ \bibnamefont {Lu}}, \bibinfo {author} {\bibfnamefont
			{R.}~\bibnamefont {Shu}}, \bibinfo {author} {\bibfnamefont {J.-Y.}\
			\bibnamefont {Wang}}, \bibinfo {author} {\bibfnamefont {L.}~\bibnamefont
			{Li}}, \bibinfo {author} {\bibfnamefont {N.}~\bibnamefont {Liu}}, \bibinfo
		{author} {\bibfnamefont {F.}~\bibnamefont {Xu}}, \bibinfo {author}
		{\bibfnamefont {X.}~\bibnamefont {Wang}}, \bibinfo {author} {\bibfnamefont
			{C.-Z.}\ \bibnamefont {Peng}},\ and\ \bibinfo {author} {\bibfnamefont
			{J.}~\bibnamefont {Pan}},\ }\bibfield  {title} {\bibinfo {title} {An
			integrated space-to-ground quantum communication network over 4,600
			kilometres},\ }\href {https://doi.org/10.1038/s41586-020-03093-8} {\bibfield
		{journal} {\bibinfo  {journal} {Nature}\ }\textbf {\bibinfo {volume} {589}},\
		\bibinfo {pages} {214} (\bibinfo {year} {2021})}\BibitemShut {NoStop}%
	\bibitem [{\citenamefont {Liu}\ \emph {et~al.}(2021)\citenamefont {Liu},
		\citenamefont {Jiang}, \citenamefont {Zhu}, \citenamefont {Zou},
		\citenamefont {Yu}, \citenamefont {Hu}, \citenamefont {Xu}, \citenamefont
		{Ma}, \citenamefont {Han}, \citenamefont {Chen}, \citenamefont {Dai},
		\citenamefont {Tang}, \citenamefont {Zhang}, \citenamefont {Li},
		\citenamefont {You}, \citenamefont {Wang}, \citenamefont {Hua}, \citenamefont
		{Hu}, \citenamefont {Zhang}, \citenamefont {Zhou}, \citenamefont {Zhang},
		\citenamefont {Wang}, \citenamefont {Chen},\ and\ \citenamefont
		{Pan}}]{PhysRevLett.126.250502}%
	\BibitemOpen
	\bibfield  {author} {\bibinfo {author} {\bibfnamefont {H.}~\bibnamefont
			{Liu}}, \bibinfo {author} {\bibfnamefont {C.}~\bibnamefont {Jiang}}, \bibinfo
		{author} {\bibfnamefont {H.-T.}\ \bibnamefont {Zhu}}, \bibinfo {author}
		{\bibfnamefont {M.}~\bibnamefont {Zou}}, \bibinfo {author} {\bibfnamefont
			{Z.-W.}\ \bibnamefont {Yu}}, \bibinfo {author} {\bibfnamefont {X.-L.}\
			\bibnamefont {Hu}}, \bibinfo {author} {\bibfnamefont {H.}~\bibnamefont {Xu}},
		\bibinfo {author} {\bibfnamefont {S.}~\bibnamefont {Ma}}, \bibinfo {author}
		{\bibfnamefont {Z.}~\bibnamefont {Han}}, \bibinfo {author} {\bibfnamefont
			{J.-P.}\ \bibnamefont {Chen}}, \bibinfo {author} {\bibfnamefont
			{Y.}~\bibnamefont {Dai}}, \bibinfo {author} {\bibfnamefont {S.-B.}\
			\bibnamefont {Tang}}, \bibinfo {author} {\bibfnamefont {W.}~\bibnamefont
			{Zhang}}, \bibinfo {author} {\bibfnamefont {H.}~\bibnamefont {Li}}, \bibinfo
		{author} {\bibfnamefont {L.}~\bibnamefont {You}}, \bibinfo {author}
		{\bibfnamefont {Z.}~\bibnamefont {Wang}}, \bibinfo {author} {\bibfnamefont
			{Y.}~\bibnamefont {Hua}}, \bibinfo {author} {\bibfnamefont {H.}~\bibnamefont
			{Hu}}, \bibinfo {author} {\bibfnamefont {H.}~\bibnamefont {Zhang}}, \bibinfo
		{author} {\bibfnamefont {F.}~\bibnamefont {Zhou}}, \bibinfo {author}
		{\bibfnamefont {Q.}~\bibnamefont {Zhang}}, \bibinfo {author} {\bibfnamefont
			{X.-B.}\ \bibnamefont {Wang}}, \bibinfo {author} {\bibfnamefont {T.-Y.}\
			\bibnamefont {Chen}},\ and\ \bibinfo {author} {\bibfnamefont {J.-W.}\
			\bibnamefont {Pan}},\ }\bibfield  {title} {\bibinfo {title} {Field test of
			twin-field quantum key distribution through sending-or-not-sending over 428
			km},\ }\href {https://doi.org/10.1103/PhysRevLett.126.250502} {\bibfield
		{journal} {\bibinfo  {journal} {Phys. Rev. Lett.}\ }\textbf {\bibinfo
			{volume} {126}},\ \bibinfo {pages} {250502} (\bibinfo {year}
		{2021})}\BibitemShut {NoStop}%
	\bibitem [{\citenamefont {Proietti}\ \emph {et~al.}(2021)\citenamefont
		{Proietti}, \citenamefont {Ho}, \citenamefont {Grasselli}, \citenamefont
		{Barrow}, \citenamefont {Malik},\ and\ \citenamefont
		{Fedrizzi}}]{proietti2021experimental}%
	\BibitemOpen
	\bibfield  {author} {\bibinfo {author} {\bibfnamefont {M.}~\bibnamefont
			{Proietti}}, \bibinfo {author} {\bibfnamefont {J.}~\bibnamefont {Ho}},
		\bibinfo {author} {\bibfnamefont {F.}~\bibnamefont {Grasselli}}, \bibinfo
		{author} {\bibfnamefont {P.}~\bibnamefont {Barrow}}, \bibinfo {author}
		{\bibfnamefont {M.}~\bibnamefont {Malik}},\ and\ \bibinfo {author}
		{\bibfnamefont {A.}~\bibnamefont {Fedrizzi}},\ }\bibfield  {title} {\bibinfo
		{title} {Experimental quantum conference key agreement},\ }\href
	{https://doi.org/10.1126/sciadv.abe0395} {\bibfield  {journal} {\bibinfo
			{journal} {Sci. Adv.}\ }\textbf {\bibinfo {volume} {7}},\ \bibinfo {pages}
		{eabe0395} (\bibinfo {year} {2021})}\BibitemShut {NoStop}%
	\bibitem [{\citenamefont {Zhong}\ \emph {et~al.}(2021)\citenamefont {Zhong},
		\citenamefont {Wang}, \citenamefont {Mandil}, \citenamefont {Lo},\ and\
		\citenamefont {Qian}}]{zhong2021experiment}%
	\BibitemOpen
	\bibfield  {author} {\bibinfo {author} {\bibfnamefont {X.}~\bibnamefont
			{Zhong}}, \bibinfo {author} {\bibfnamefont {W.}~\bibnamefont {Wang}},
		\bibinfo {author} {\bibfnamefont {R.}~\bibnamefont {Mandil}}, \bibinfo
		{author} {\bibfnamefont {H.-K.}\ \bibnamefont {Lo}},\ and\ \bibinfo {author}
		{\bibfnamefont {L.}~\bibnamefont {Qian}},\ }\bibfield  {title} {\bibinfo
		{title} {Experiment on scalable multi-user twin-field quantum key
			distribution network},\ }\href@noop {} {\bibfield  {journal} {\bibinfo
			{journal} {arXiv preprint arXiv:2106.07768}\ } (\bibinfo {year}
		{2021})}\BibitemShut {NoStop}%
	\bibitem [{\citenamefont {Cerf}\ \emph {et~al.}(2002)\citenamefont {Cerf},
		\citenamefont {Bourennane}, \citenamefont {Karlsson},\ and\ \citenamefont
		{Gisin}}]{PhysRevLett.88.127902}%
	\BibitemOpen
	\bibfield  {author} {\bibinfo {author} {\bibfnamefont {N.~J.}\ \bibnamefont
			{Cerf}}, \bibinfo {author} {\bibfnamefont {M.}~\bibnamefont {Bourennane}},
		\bibinfo {author} {\bibfnamefont {A.}~\bibnamefont {Karlsson}},\ and\
		\bibinfo {author} {\bibfnamefont {N.}~\bibnamefont {Gisin}},\ }\bibfield
	{title} {\bibinfo {title} {Security of quantum key distribution using
			$\mathit{d}$-level systems},\ }\href
	{https://doi.org/10.1103/PhysRevLett.88.127902} {\bibfield  {journal}
		{\bibinfo  {journal} {Phys. Rev. Lett.}\ }\textbf {\bibinfo {volume} {88}},\
		\bibinfo {pages} {127902} (\bibinfo {year} {2002})}\BibitemShut {NoStop}%
	\bibitem [{\citenamefont {Vaziri}\ \emph {et~al.}(2002)\citenamefont {Vaziri},
		\citenamefont {Weihs},\ and\ \citenamefont
		{Zeilinger}}]{PhysRevLett.89.240401}%
	\BibitemOpen
	\bibfield  {author} {\bibinfo {author} {\bibfnamefont {A.}~\bibnamefont
			{Vaziri}}, \bibinfo {author} {\bibfnamefont {G.}~\bibnamefont {Weihs}},\ and\
		\bibinfo {author} {\bibfnamefont {A.}~\bibnamefont {Zeilinger}},\ }\bibfield
	{title} {\bibinfo {title} {Experimental two-photon, three-dimensional
			entanglement for quantum communication},\ }\href
	{https://doi.org/10.1103/PhysRevLett.89.240401} {\bibfield  {journal}
		{\bibinfo  {journal} {Phys. Rev. Lett.}\ }\textbf {\bibinfo {volume} {89}},\
		\bibinfo {pages} {240401} (\bibinfo {year} {2002})}\BibitemShut {NoStop}%
	\bibitem [{\citenamefont {Gröblacher}\ \emph {et~al.}(2006)\citenamefont
		{Gröblacher}, \citenamefont {Jennewein}, \citenamefont {Vaziri},
		\citenamefont {Weihs},\ and\ \citenamefont {Zeilinger}}]{Gr_blacher_2006}%
	\BibitemOpen
	\bibfield  {author} {\bibinfo {author} {\bibfnamefont {S.}~\bibnamefont
			{Gröblacher}}, \bibinfo {author} {\bibfnamefont {T.}~\bibnamefont
			{Jennewein}}, \bibinfo {author} {\bibfnamefont {A.}~\bibnamefont {Vaziri}},
		\bibinfo {author} {\bibfnamefont {G.}~\bibnamefont {Weihs}},\ and\ \bibinfo
		{author} {\bibfnamefont {A.}~\bibnamefont {Zeilinger}},\ }\bibfield  {title}
	{\bibinfo {title} {Experimental quantum cryptography with qutrits},\ }\href
	{https://doi.org/10.1088/1367-2630/8/5/075} {\bibfield  {journal} {\bibinfo
			{journal} {New J. Phys.}\ }\textbf {\bibinfo {volume} {8}},\ \bibinfo {pages}
		{75} (\bibinfo {year} {2006})}\BibitemShut {NoStop}%
	\bibitem [{\citenamefont {Mafu}\ \emph {et~al.}(2013)\citenamefont {Mafu},
		\citenamefont {Dudley}, \citenamefont {Goyal}, \citenamefont {Giovannini},
		\citenamefont {McLaren}, \citenamefont {Padgett}, \citenamefont {Konrad},
		\citenamefont {Petruccione}, \citenamefont {L\"utkenhaus},\ and\
		\citenamefont {Forbes}}]{PhysRevA.88.032305}%
	\BibitemOpen
	\bibfield  {author} {\bibinfo {author} {\bibfnamefont {M.}~\bibnamefont
			{Mafu}}, \bibinfo {author} {\bibfnamefont {A.}~\bibnamefont {Dudley}},
		\bibinfo {author} {\bibfnamefont {S.}~\bibnamefont {Goyal}}, \bibinfo
		{author} {\bibfnamefont {D.}~\bibnamefont {Giovannini}}, \bibinfo {author}
		{\bibfnamefont {M.}~\bibnamefont {McLaren}}, \bibinfo {author} {\bibfnamefont
			{M.~J.}\ \bibnamefont {Padgett}}, \bibinfo {author} {\bibfnamefont
			{T.}~\bibnamefont {Konrad}}, \bibinfo {author} {\bibfnamefont
			{F.}~\bibnamefont {Petruccione}}, \bibinfo {author} {\bibfnamefont
			{N.}~\bibnamefont {L\"utkenhaus}},\ and\ \bibinfo {author} {\bibfnamefont
			{A.}~\bibnamefont {Forbes}},\ }\bibfield  {title} {\bibinfo {title}
		{Higher-dimensional orbital-angular-momentum-based quantum key distribution
			with mutually unbiased bases},\ }\href
	{https://doi.org/10.1103/PhysRevA.88.032305} {\bibfield  {journal} {\bibinfo
			{journal} {Phys. Rev. A}\ }\textbf {\bibinfo {volume} {88}},\ \bibinfo
		{pages} {032305} (\bibinfo {year} {2013})}\BibitemShut {NoStop}%
	\bibitem [{\citenamefont {Mirhosseini}\ \emph {et~al.}(2015)\citenamefont
		{Mirhosseini}, \citenamefont {Maga{\~{n}}a-Loaiza}, \citenamefont
		{O'Sullivan}, \citenamefont {Rodenburg}, \citenamefont {Malik}, \citenamefont
		{Lavery}, \citenamefont {Padgett}, \citenamefont {Gauthier},\ and\
		\citenamefont {Boyd}}]{Mirhosseini_2015}%
	\BibitemOpen
	\bibfield  {author} {\bibinfo {author} {\bibfnamefont {M.}~\bibnamefont
			{Mirhosseini}}, \bibinfo {author} {\bibfnamefont {O.~S.}\ \bibnamefont
			{Maga{\~{n}}a-Loaiza}}, \bibinfo {author} {\bibfnamefont {M.~N.}\
			\bibnamefont {O'Sullivan}}, \bibinfo {author} {\bibfnamefont
			{B.}~\bibnamefont {Rodenburg}}, \bibinfo {author} {\bibfnamefont
			{M.}~\bibnamefont {Malik}}, \bibinfo {author} {\bibfnamefont {M.~P.~J.}\
			\bibnamefont {Lavery}}, \bibinfo {author} {\bibfnamefont {M.~J.}\
			\bibnamefont {Padgett}}, \bibinfo {author} {\bibfnamefont {D.~J.}\
			\bibnamefont {Gauthier}},\ and\ \bibinfo {author} {\bibfnamefont {R.~W.}\
			\bibnamefont {Boyd}},\ }\bibfield  {title} {\bibinfo {title}
		{High-dimensional quantum cryptography with twisted light},\ }\href
	{https://doi.org/10.1088/1367-2630/17/3/033033} {\bibfield  {journal}
		{\bibinfo  {journal} {New J. Phys.}\ }\textbf {\bibinfo {volume} {17}},\
		\bibinfo {pages} {033033} (\bibinfo {year} {2015})}\BibitemShut {NoStop}%
	\bibitem [{\citenamefont {Lucamarini}\ \emph {et~al.}(2018)\citenamefont
		{Lucamarini}, \citenamefont {Yuan}, \citenamefont {Dynes},\ and\
		\citenamefont {Shields}}]{TF-QKD}%
	\BibitemOpen
	\bibfield  {author} {\bibinfo {author} {\bibfnamefont {M.}~\bibnamefont
			{Lucamarini}}, \bibinfo {author} {\bibfnamefont {Z.~L.}\ \bibnamefont
			{Yuan}}, \bibinfo {author} {\bibfnamefont {J.~F.}\ \bibnamefont {Dynes}},\
		and\ \bibinfo {author} {\bibfnamefont {A.~J.}\ \bibnamefont {Shields}},\
	}\bibfield  {title} {\bibinfo {title} {Overcoming the rate–distance limit
			of quantum key distribution without quantum repeaters},\ }\href
	{https://doi.org/10.1038/s41586-018-0066-6} {\bibfield  {journal} {\bibinfo
			{journal} {Nature}\ }\textbf {\bibinfo {volume} {557}},\ \bibinfo {pages}
		{400–403} (\bibinfo {year} {2018})}\BibitemShut {NoStop}%
	\bibitem [{\citenamefont {Ma}\ \emph {et~al.}(2018)\citenamefont {Ma},
		\citenamefont {Zeng},\ and\ \citenamefont {Zhou}}]{PhysRevX.8.031043}%
	\BibitemOpen
	\bibfield  {author} {\bibinfo {author} {\bibfnamefont {X.}~\bibnamefont
			{Ma}}, \bibinfo {author} {\bibfnamefont {P.}~\bibnamefont {Zeng}},\ and\
		\bibinfo {author} {\bibfnamefont {H.}~\bibnamefont {Zhou}},\ }\bibfield
	{title} {\bibinfo {title} {Phase-matching quantum key distribution},\ }\href
	{https://doi.org/10.1103/PhysRevX.8.031043} {\bibfield  {journal} {\bibinfo
			{journal} {Phys. Rev. X}\ }\textbf {\bibinfo {volume} {8}},\ \bibinfo {pages}
		{031043} (\bibinfo {year} {2018})}\BibitemShut {NoStop}%
	\bibitem [{\citenamefont {Wang}\ \emph {et~al.}(2018)\citenamefont {Wang},
		\citenamefont {Yu},\ and\ \citenamefont {Hu}}]{PhysRevA.98.062323}%
	\BibitemOpen
	\bibfield  {author} {\bibinfo {author} {\bibfnamefont {X.-B.}\ \bibnamefont
			{Wang}}, \bibinfo {author} {\bibfnamefont {Z.-W.}\ \bibnamefont {Yu}},\ and\
		\bibinfo {author} {\bibfnamefont {X.-L.}\ \bibnamefont {Hu}},\ }\bibfield
	{title} {\bibinfo {title} {Twin-field quantum key distribution with large
			misalignment error},\ }\href {https://doi.org/10.1103/PhysRevA.98.062323}
	{\bibfield  {journal} {\bibinfo  {journal} {Phys. Rev. A}\ }\textbf {\bibinfo
			{volume} {98}},\ \bibinfo {pages} {062323} (\bibinfo {year}
		{2018})}\BibitemShut {NoStop}%
	\bibitem [{\citenamefont {Gisin}\ \emph {et~al.}(2002)\citenamefont {Gisin},
		\citenamefont {Ribordy}, \citenamefont {Tittel},\ and\ \citenamefont
		{Zbinden}}]{RevModPhys.74.145}%
	\BibitemOpen
	\bibfield  {author} {\bibinfo {author} {\bibfnamefont {N.}~\bibnamefont
			{Gisin}}, \bibinfo {author} {\bibfnamefont {G.}~\bibnamefont {Ribordy}},
		\bibinfo {author} {\bibfnamefont {W.}~\bibnamefont {Tittel}},\ and\ \bibinfo
		{author} {\bibfnamefont {H.}~\bibnamefont {Zbinden}},\ }\bibfield  {title}
	{\bibinfo {title} {Quantum cryptography},\ }\href
	{https://doi.org/10.1103/RevModPhys.74.145} {\bibfield  {journal} {\bibinfo
			{journal} {Rev. Mod. Phys.}\ }\textbf {\bibinfo {volume} {74}},\ \bibinfo
		{pages} {145} (\bibinfo {year} {2002})}\BibitemShut {NoStop}%
	\bibitem [{\citenamefont {Scarani}\ \emph {et~al.}(2009)\citenamefont
		{Scarani}, \citenamefont {Bechmann-Pasquinucci}, \citenamefont {Cerf},
		\citenamefont {Du\ifmmode~\check{s}\else \v{s}\fi{}ek}, \citenamefont
		{L\"utkenhaus},\ and\ \citenamefont {Peev}}]{RevModPhys.81.1301}%
	\BibitemOpen
	\bibfield  {author} {\bibinfo {author} {\bibfnamefont {V.}~\bibnamefont
			{Scarani}}, \bibinfo {author} {\bibfnamefont {H.}~\bibnamefont
			{Bechmann-Pasquinucci}}, \bibinfo {author} {\bibfnamefont {N.~J.}\
			\bibnamefont {Cerf}}, \bibinfo {author} {\bibfnamefont {M.}~\bibnamefont
			{Du\ifmmode~\check{s}\else \v{s}\fi{}ek}}, \bibinfo {author} {\bibfnamefont
			{N.}~\bibnamefont {L\"utkenhaus}},\ and\ \bibinfo {author} {\bibfnamefont
			{M.}~\bibnamefont {Peev}},\ }\bibfield  {title} {\bibinfo {title} {The
			security of practical quantum key distribution},\ }\href
	{https://doi.org/10.1103/RevModPhys.81.1301} {\bibfield  {journal} {\bibinfo
			{journal} {Rev. Mod. Phys.}\ }\textbf {\bibinfo {volume} {81}},\ \bibinfo
		{pages} {1301} (\bibinfo {year} {2009})}\BibitemShut {NoStop}%
	\bibitem [{\citenamefont {Xu}\ \emph {et~al.}(2020)\citenamefont {Xu},
		\citenamefont {Ma}, \citenamefont {Zhang}, \citenamefont {Lo},\ and\
		\citenamefont {Pan}}]{RevModPhys.92.025002}%
	\BibitemOpen
	\bibfield  {author} {\bibinfo {author} {\bibfnamefont {F.}~\bibnamefont
			{Xu}}, \bibinfo {author} {\bibfnamefont {X.}~\bibnamefont {Ma}}, \bibinfo
		{author} {\bibfnamefont {Q.}~\bibnamefont {Zhang}}, \bibinfo {author}
		{\bibfnamefont {H.-K.}\ \bibnamefont {Lo}},\ and\ \bibinfo {author}
		{\bibfnamefont {J.-W.}\ \bibnamefont {Pan}},\ }\bibfield  {title} {\bibinfo
		{title} {Secure quantum key distribution with realistic devices},\ }\href
	{https://doi.org/10.1103/RevModPhys.92.025002} {\bibfield  {journal}
		{\bibinfo  {journal} {Rev. Mod. Phys.}\ }\textbf {\bibinfo {volume} {92}},\
		\bibinfo {pages} {025002} (\bibinfo {year} {2020})}\BibitemShut {NoStop}%
	\bibitem [{\citenamefont {Gottesman}\ \emph {et~al.}(2004)\citenamefont
		{Gottesman}, \citenamefont {Lo}, \citenamefont {L{\"u}tkenhaus},\ and\
		\citenamefont {Preskill}}]{1365172}%
	\BibitemOpen
	\bibfield  {author} {\bibinfo {author} {\bibfnamefont {D.}~\bibnamefont
			{Gottesman}}, \bibinfo {author} {\bibfnamefont {H.-K.}\ \bibnamefont {Lo}},
		\bibinfo {author} {\bibfnamefont {N.}~\bibnamefont {L{\"u}tkenhaus}},\ and\
		\bibinfo {author} {\bibfnamefont {J.}~\bibnamefont {Preskill}},\ }\bibfield
	{title} {\bibinfo {title} {Security of quantum key distribution with
			imperfect devices},\ }\href {https://dl.acm.org/doi/10.5555/2011586.2011587}
	{\bibfield  {journal} {\bibinfo  {journal} {Quantum Inf Comput}\ }\textbf
		{\bibinfo {volume} {4}},\ \bibinfo {pages} {325} (\bibinfo {year}
		{2004})}\BibitemShut {NoStop}%
	\bibitem [{\citenamefont {Jain}\ \emph {et~al.}(2011)\citenamefont {Jain},
		\citenamefont {Wittmann}, \citenamefont {Lydersen}, \citenamefont {Wiechers},
		\citenamefont {Elser}, \citenamefont {Marquardt}, \citenamefont {Makarov},\
		and\ \citenamefont {Leuchs}}]{PhysRevLett.107.110501}%
	\BibitemOpen
	\bibfield  {author} {\bibinfo {author} {\bibfnamefont {N.}~\bibnamefont
			{Jain}}, \bibinfo {author} {\bibfnamefont {C.}~\bibnamefont {Wittmann}},
		\bibinfo {author} {\bibfnamefont {L.}~\bibnamefont {Lydersen}}, \bibinfo
		{author} {\bibfnamefont {C.}~\bibnamefont {Wiechers}}, \bibinfo {author}
		{\bibfnamefont {D.}~\bibnamefont {Elser}}, \bibinfo {author} {\bibfnamefont
			{C.}~\bibnamefont {Marquardt}}, \bibinfo {author} {\bibfnamefont
			{V.}~\bibnamefont {Makarov}},\ and\ \bibinfo {author} {\bibfnamefont
			{G.}~\bibnamefont {Leuchs}},\ }\bibfield  {title} {\bibinfo {title} {Device
			calibration impacts security of quantum key distribution},\ }\href
	{https://doi.org/10.1103/PhysRevLett.107.110501} {\bibfield  {journal}
		{\bibinfo  {journal} {Phys. Rev. Lett.}\ }\textbf {\bibinfo {volume} {107}},\
		\bibinfo {pages} {110501} (\bibinfo {year} {2011})}\BibitemShut {NoStop}%
	\bibitem [{\citenamefont {Dong}\ \emph {et~al.}(2014)\citenamefont {Dong},
		\citenamefont {Yu}, \citenamefont {Wei}, \citenamefont {Wang},\ and\
		\citenamefont {Zhang}}]{dong2014attack}%
	\BibitemOpen
	\bibfield  {author} {\bibinfo {author} {\bibfnamefont {Z.-Y.}\ \bibnamefont
			{Dong}}, \bibinfo {author} {\bibfnamefont {N.-N.}\ \bibnamefont {Yu}},
		\bibinfo {author} {\bibfnamefont {Z.-J.}\ \bibnamefont {Wei}}, \bibinfo
		{author} {\bibfnamefont {J.-D.}\ \bibnamefont {Wang}},\ and\ \bibinfo
		{author} {\bibfnamefont {Z.-M.}\ \bibnamefont {Zhang}},\ }\bibfield  {title}
	{\bibinfo {title} {An attack aimed at active phase compensation in one-way
			phase-encoded qkd systems},\ }\href
	{https://doi.org/10.1140/epjd/e2014-40693-6} {\bibfield  {journal} {\bibinfo
			{journal} {Eur. Phys. J. D}\ }\textbf {\bibinfo {volume} {68}},\ \bibinfo
		{pages} {1} (\bibinfo {year} {2014})}\BibitemShut {NoStop}%
	\bibitem [{\citenamefont {Paterson}(2005)}]{PhysRevLett.94.153901}%
	\BibitemOpen
	\bibfield  {author} {\bibinfo {author} {\bibfnamefont {C.}~\bibnamefont
			{Paterson}},\ }\bibfield  {title} {\bibinfo {title} {Atmospheric turbulence
			and orbital angular momentum of single photons for optical communication},\
	}\href {https://doi.org/10.1103/PhysRevLett.94.153901} {\bibfield  {journal}
		{\bibinfo  {journal} {Phys. Rev. Lett.}\ }\textbf {\bibinfo {volume} {94}},\
		\bibinfo {pages} {153901} (\bibinfo {year} {2005})}\BibitemShut {NoStop}%
	\bibitem [{\citenamefont {Tamaki}\ \emph {et~al.}(2014)\citenamefont {Tamaki},
		\citenamefont {Curty}, \citenamefont {Kato}, \citenamefont {Lo},\ and\
		\citenamefont {Azuma}}]{tamaki2014loss}%
	\BibitemOpen
	\bibfield  {author} {\bibinfo {author} {\bibfnamefont {K.}~\bibnamefont
			{Tamaki}}, \bibinfo {author} {\bibfnamefont {M.}~\bibnamefont {Curty}},
		\bibinfo {author} {\bibfnamefont {G.}~\bibnamefont {Kato}}, \bibinfo {author}
		{\bibfnamefont {H.-K.}\ \bibnamefont {Lo}},\ and\ \bibinfo {author}
		{\bibfnamefont {K.}~\bibnamefont {Azuma}},\ }\bibfield  {title} {\bibinfo
		{title} {Loss-tolerant quantum cryptography with imperfect sources},\ }\href
	{https://doi.org/10.1103/PhysRevA.90.052314} {\bibfield  {journal} {\bibinfo
			{journal} {Phys. Rev. A}\ }\textbf {\bibinfo {volume} {90}},\ \bibinfo
		{pages} {052314} (\bibinfo {year} {2014})}\BibitemShut {NoStop}%
	\bibitem [{\citenamefont {Yin}\ \emph {et~al.}(2014)\citenamefont {Yin},
		\citenamefont {Fung}, \citenamefont {Ma}, \citenamefont {Zhang},
		\citenamefont {Li}, \citenamefont {Chen}, \citenamefont {Wang}, \citenamefont
		{Guo},\ and\ \citenamefont {Han}}]{PhysRevA.90.052319}%
	\BibitemOpen
	\bibfield  {author} {\bibinfo {author} {\bibfnamefont {Z.-Q.}\ \bibnamefont
			{Yin}}, \bibinfo {author} {\bibfnamefont {C.-H.~F.}\ \bibnamefont {Fung}},
		\bibinfo {author} {\bibfnamefont {X.}~\bibnamefont {Ma}}, \bibinfo {author}
		{\bibfnamefont {C.-M.}\ \bibnamefont {Zhang}}, \bibinfo {author}
		{\bibfnamefont {H.-W.}\ \bibnamefont {Li}}, \bibinfo {author} {\bibfnamefont
			{W.}~\bibnamefont {Chen}}, \bibinfo {author} {\bibfnamefont {S.}~\bibnamefont
			{Wang}}, \bibinfo {author} {\bibfnamefont {G.-C.}\ \bibnamefont {Guo}},\ and\
		\bibinfo {author} {\bibfnamefont {Z.-F.}\ \bibnamefont {Han}},\ }\bibfield
	{title} {\bibinfo {title} {Mismatched-basis statistics enable quantum key
			distribution with uncharacterized qubit sources},\ }\href
	{https://doi.org/10.1103/PhysRevA.90.052319} {\bibfield  {journal} {\bibinfo
			{journal} {Phys. Rev. A}\ }\textbf {\bibinfo {volume} {90}},\ \bibinfo
		{pages} {052319} (\bibinfo {year} {2014})}\BibitemShut {NoStop}%
	\bibitem [{\citenamefont {Coles}\ \emph {et~al.}(2016)\citenamefont {Coles},
		\citenamefont {Metodiev},\ and\ \citenamefont {Lutkenhaus}}]{Coles}%
	\BibitemOpen
	\bibfield  {author} {\bibinfo {author} {\bibfnamefont {P.~J.}\ \bibnamefont
			{Coles}}, \bibinfo {author} {\bibfnamefont {E.~M.}\ \bibnamefont
			{Metodiev}},\ and\ \bibinfo {author} {\bibfnamefont {N.}~\bibnamefont
			{Lutkenhaus}},\ }\bibfield  {title} {\bibinfo {title} {Numerical approach for
			unstructured quantum key distribution},\ }\href
	{https://doi.org/10.1038/ncomms11712} {\bibfield  {journal} {\bibinfo
			{journal} {Nat. Commun.}\ }\textbf {\bibinfo {volume} {7}},\ \bibinfo {pages}
		{11712} (\bibinfo {year} {2016})}\BibitemShut {NoStop}%
	\bibitem [{\citenamefont {Bennett}(1992)}]{PhysRevLett.68.3121}%
	\BibitemOpen
	\bibfield  {author} {\bibinfo {author} {\bibfnamefont {C.~H.}\ \bibnamefont
			{Bennett}},\ }\bibfield  {title} {\bibinfo {title} {Quantum cryptography
			using any two nonorthogonal states},\ }\href
	{https://doi.org/10.1103/PhysRevLett.68.3121} {\bibfield  {journal} {\bibinfo
			{journal} {Phys. Rev. Lett.}\ }\textbf {\bibinfo {volume} {68}},\ \bibinfo
		{pages} {3121} (\bibinfo {year} {1992})}\BibitemShut {NoStop}%
	\bibitem [{\citenamefont {Matsumoto}\ and\ \citenamefont
		{Watanabe}(2010)}]{Matsumoto_2010}%
	\BibitemOpen
	\bibfield  {author} {\bibinfo {author} {\bibfnamefont {R.}~\bibnamefont
			{Matsumoto}}\ and\ \bibinfo {author} {\bibfnamefont {S.}~\bibnamefont
			{Watanabe}},\ }\bibfield  {title} {\bibinfo {title} {Narrow basis angle
			doubles secret key in the {BB}84 protocol},\ }\href
	{https://doi.org/10.1088/1751-8113/43/14/145302} {\bibfield  {journal}
		{\bibinfo  {journal} {J. Phys. A Math. Theor.}\ }\textbf {\bibinfo {volume}
			{43}},\ \bibinfo {pages} {145302} (\bibinfo {year} {2010})}\BibitemShut
	{NoStop}%
	\bibitem [{\citenamefont {Liu}\ \emph {et~al.}(2019)\citenamefont {Liu},
		\citenamefont {Wang}, \citenamefont {Ma},\ and\ \citenamefont
		{Sun}}]{PhysRevApplied.12.034039}%
	\BibitemOpen
	\bibfield  {author} {\bibinfo {author} {\bibfnamefont {H.}~\bibnamefont
			{Liu}}, \bibinfo {author} {\bibfnamefont {J.}~\bibnamefont {Wang}}, \bibinfo
		{author} {\bibfnamefont {H.}~\bibnamefont {Ma}},\ and\ \bibinfo {author}
		{\bibfnamefont {S.}~\bibnamefont {Sun}},\ }\bibfield  {title} {\bibinfo
		{title} {Reference-frame-independent quantum key distribution using fewer
			states},\ }\href {https://doi.org/10.1103/PhysRevApplied.12.034039}
	{\bibfield  {journal} {\bibinfo  {journal} {Phys. Rev. Appl.}\ }\textbf
		{\bibinfo {volume} {12}},\ \bibinfo {pages} {034039} (\bibinfo {year}
		{2019})}\BibitemShut {NoStop}%
	\bibitem [{\citenamefont {Laing}\ \emph {et~al.}(2010)\citenamefont {Laing},
		\citenamefont {Scarani}, \citenamefont {Rarity},\ and\ \citenamefont
		{O'Brien}}]{PhysRevA.82.012304}%
	\BibitemOpen
	\bibfield  {author} {\bibinfo {author} {\bibfnamefont {A.}~\bibnamefont
			{Laing}}, \bibinfo {author} {\bibfnamefont {V.}~\bibnamefont {Scarani}},
		\bibinfo {author} {\bibfnamefont {J.~G.}\ \bibnamefont {Rarity}},\ and\
		\bibinfo {author} {\bibfnamefont {J.~L.}\ \bibnamefont {O'Brien}},\
	}\bibfield  {title} {\bibinfo {title} {Reference-frame-independent quantum
			key distribution},\ }\href {https://doi.org/10.1103/PhysRevA.82.012304}
	{\bibfield  {journal} {\bibinfo  {journal} {Phys. Rev. A}\ }\textbf {\bibinfo
			{volume} {82}},\ \bibinfo {pages} {012304} (\bibinfo {year}
		{2010})}\BibitemShut {NoStop}%
	\bibitem [{\citenamefont {Bunandar}\ \emph {et~al.}(2020)\citenamefont
		{Bunandar}, \citenamefont {Govia}, \citenamefont {Krovi},\ and\ \citenamefont
		{Englund}}]{numericalnpj}%
	\BibitemOpen
	\bibfield  {author} {\bibinfo {author} {\bibfnamefont {D.}~\bibnamefont
			{Bunandar}}, \bibinfo {author} {\bibfnamefont {L.~C.~G.}\ \bibnamefont
			{Govia}}, \bibinfo {author} {\bibfnamefont {H.}~\bibnamefont {Krovi}},\ and\
		\bibinfo {author} {\bibfnamefont {D.}~\bibnamefont {Englund}},\ }\bibfield
	{title} {\bibinfo {title} {Numerical finite-key analysis of quantum key
			distribution},\ }\href {https://doi.org/10.1038/s41534-020-00322-w}
	{\bibfield  {journal} {\bibinfo  {journal} {NPJ Quantum Inf.}\ }\textbf
		{\bibinfo {volume} {6}},\ \bibinfo {pages} {104} (\bibinfo {year}
		{2020})}\BibitemShut {NoStop}%
	\bibitem [{\citenamefont {George}\ \emph {et~al.}(2021)\citenamefont {George},
		\citenamefont {Lin},\ and\ \citenamefont
		{L\"utkenhaus}}]{PhysRevResearch.3.013274}%
	\BibitemOpen
	\bibfield  {author} {\bibinfo {author} {\bibfnamefont {I.}~\bibnamefont
			{George}}, \bibinfo {author} {\bibfnamefont {J.}~\bibnamefont {Lin}},\ and\
		\bibinfo {author} {\bibfnamefont {N.}~\bibnamefont {L\"utkenhaus}},\
	}\bibfield  {title} {\bibinfo {title} {Numerical calculations of the finite
			key rate for general quantum key distribution protocols},\ }\href
	{https://doi.org/10.1103/PhysRevResearch.3.013274} {\bibfield  {journal}
		{\bibinfo  {journal} {Phys. Rev. Research}\ }\textbf {\bibinfo {volume}
			{3}},\ \bibinfo {pages} {013274} (\bibinfo {year} {2021})}\BibitemShut
	{NoStop}%
	\bibitem [{\citenamefont {Wang}\ \emph {et~al.}(2020)\citenamefont {Wang},
		\citenamefont {Zeng}, \citenamefont {Zhao}, \citenamefont {Braverman},
		\citenamefont {Zhou}, \citenamefont {Mirhosseini}, \citenamefont {Wang},
		\citenamefont {Gao}, \citenamefont {Li}, \citenamefont {Boyd},\ and\
		\citenamefont {Zhang}}]{PhysRevA.101.032340}%
	\BibitemOpen
	\bibfield  {author} {\bibinfo {author} {\bibfnamefont {F.}~\bibnamefont
			{Wang}}, \bibinfo {author} {\bibfnamefont {P.}~\bibnamefont {Zeng}}, \bibinfo
		{author} {\bibfnamefont {J.}~\bibnamefont {Zhao}}, \bibinfo {author}
		{\bibfnamefont {B.}~\bibnamefont {Braverman}}, \bibinfo {author}
		{\bibfnamefont {Y.}~\bibnamefont {Zhou}}, \bibinfo {author} {\bibfnamefont
			{M.}~\bibnamefont {Mirhosseini}}, \bibinfo {author} {\bibfnamefont
			{X.}~\bibnamefont {Wang}}, \bibinfo {author} {\bibfnamefont {H.}~\bibnamefont
			{Gao}}, \bibinfo {author} {\bibfnamefont {F.}~\bibnamefont {Li}}, \bibinfo
		{author} {\bibfnamefont {R.~W.}\ \bibnamefont {Boyd}},\ and\ \bibinfo
		{author} {\bibfnamefont {P.}~\bibnamefont {Zhang}},\ }\bibfield  {title}
	{\bibinfo {title} {High-dimensional quantum key distribution based on
			mutually partially unbiased bases},\ }\href
	{https://doi.org/10.1103/PhysRevA.101.032340} {\bibfield  {journal} {\bibinfo
			{journal} {Phys. Rev. A}\ }\textbf {\bibinfo {volume} {101}},\ \bibinfo
		{pages} {032340} (\bibinfo {year} {2020})}\BibitemShut {NoStop}%
	\bibitem [{\citenamefont {Koashi}(2007)}]{koashi2007complementarity}%
	\BibitemOpen
	\bibfield  {author} {\bibinfo {author} {\bibfnamefont {M.}~\bibnamefont
			{Koashi}},\ }\bibfield  {title} {\bibinfo {title} {Complementarity,
			distillable secret key, and distillable entanglement},\ }\href@noop {}
	{\bibfield  {journal} {\bibinfo  {journal} {arXiv preprint arXiv:0704.3661}\
		} (\bibinfo {year} {2007})}\BibitemShut {NoStop}%
	\bibitem [{\citenamefont {Zhang}\ \emph {et~al.}(2014)\citenamefont {Zhang},
		\citenamefont {Aungskunsiri}, \citenamefont {Mart\'{\i}n-L\'opez},
		\citenamefont {Wabnig}, \citenamefont {Lobino}, \citenamefont {Nock},
		\citenamefont {Munns}, \citenamefont {Bonneau}, \citenamefont {Jiang},
		\citenamefont {Li}, \citenamefont {Laing}, \citenamefont {Rarity},
		\citenamefont {Niskanen}, \citenamefont {Thompson},\ and\ \citenamefont
		{O'Brien}}]{PhysRevLett.112.130501}%
	\BibitemOpen
	\bibfield  {author} {\bibinfo {author} {\bibfnamefont {P.}~\bibnamefont
			{Zhang}}, \bibinfo {author} {\bibfnamefont {K.}~\bibnamefont {Aungskunsiri}},
		\bibinfo {author} {\bibfnamefont {E.}~\bibnamefont {Mart\'{\i}n-L\'opez}},
		\bibinfo {author} {\bibfnamefont {J.}~\bibnamefont {Wabnig}}, \bibinfo
		{author} {\bibfnamefont {M.}~\bibnamefont {Lobino}}, \bibinfo {author}
		{\bibfnamefont {R.~W.}\ \bibnamefont {Nock}}, \bibinfo {author}
		{\bibfnamefont {J.}~\bibnamefont {Munns}}, \bibinfo {author} {\bibfnamefont
			{D.}~\bibnamefont {Bonneau}}, \bibinfo {author} {\bibfnamefont
			{P.}~\bibnamefont {Jiang}}, \bibinfo {author} {\bibfnamefont {H.~W.}\
			\bibnamefont {Li}}, \bibinfo {author} {\bibfnamefont {A.}~\bibnamefont
			{Laing}}, \bibinfo {author} {\bibfnamefont {J.~G.}\ \bibnamefont {Rarity}},
		\bibinfo {author} {\bibfnamefont {A.~O.}\ \bibnamefont {Niskanen}}, \bibinfo
		{author} {\bibfnamefont {M.~G.}\ \bibnamefont {Thompson}},\ and\ \bibinfo
		{author} {\bibfnamefont {J.~L.}\ \bibnamefont {O'Brien}},\ }\bibfield
	{title} {\bibinfo {title} {Reference-frame-independent
			quantum-key-distribution server with a telecom tether for an on-chip
			client},\ }\href {https://doi.org/10.1103/PhysRevLett.112.130501} {\bibfield
		{journal} {\bibinfo  {journal} {Phys. Rev. Lett.}\ }\textbf {\bibinfo
			{volume} {112}},\ \bibinfo {pages} {130501} (\bibinfo {year}
		{2014})}\BibitemShut {NoStop}%
	\bibitem [{\citenamefont {Xu}\ \emph {et~al.}(2015)\citenamefont {Xu},
		\citenamefont {Wei}, \citenamefont {Sajeed}, \citenamefont {Kaiser},
		\citenamefont {Sun}, \citenamefont {Tang}, \citenamefont {Qian},
		\citenamefont {Makarov},\ and\ \citenamefont {Lo}}]{xu2015experimental}%
	\BibitemOpen
	\bibfield  {author} {\bibinfo {author} {\bibfnamefont {F.}~\bibnamefont
			{Xu}}, \bibinfo {author} {\bibfnamefont {K.}~\bibnamefont {Wei}}, \bibinfo
		{author} {\bibfnamefont {S.}~\bibnamefont {Sajeed}}, \bibinfo {author}
		{\bibfnamefont {S.}~\bibnamefont {Kaiser}}, \bibinfo {author} {\bibfnamefont
			{S.}~\bibnamefont {Sun}}, \bibinfo {author} {\bibfnamefont {Z.}~\bibnamefont
			{Tang}}, \bibinfo {author} {\bibfnamefont {L.}~\bibnamefont {Qian}}, \bibinfo
		{author} {\bibfnamefont {V.}~\bibnamefont {Makarov}},\ and\ \bibinfo {author}
		{\bibfnamefont {H.-K.}\ \bibnamefont {Lo}},\ }\bibfield  {title} {\bibinfo
		{title} {Experimental quantum key distribution with source flaws},\ }\href
	{https://doi.org/10.1103/PhysRevA.92.032305} {\bibfield  {journal} {\bibinfo
			{journal} {Phys. Rev. A}\ }\textbf {\bibinfo {volume} {92}},\ \bibinfo
		{pages} {032305} (\bibinfo {year} {2015})}\BibitemShut {NoStop}%
	\bibitem [{\citenamefont {Ma}\ \emph {et~al.}(2005)\citenamefont {Ma},
		\citenamefont {Qi}, \citenamefont {Zhao},\ and\ \citenamefont
		{Lo}}]{ma2005practical}%
	\BibitemOpen
	\bibfield  {author} {\bibinfo {author} {\bibfnamefont {X.}~\bibnamefont
			{Ma}}, \bibinfo {author} {\bibfnamefont {B.}~\bibnamefont {Qi}}, \bibinfo
		{author} {\bibfnamefont {Y.}~\bibnamefont {Zhao}},\ and\ \bibinfo {author}
		{\bibfnamefont {H.-K.}\ \bibnamefont {Lo}},\ }\bibfield  {title} {\bibinfo
		{title} {Practical decoy state for quantum key distribution},\ }\href
	{https://doi.org/10.1103/PhysRevA.72.012326} {\bibfield  {journal} {\bibinfo
			{journal} {Phys. Rev. A}\ }\textbf {\bibinfo {volume} {72}},\ \bibinfo
		{pages} {012326} (\bibinfo {year} {2005})}\BibitemShut {NoStop}%
	\bibitem [{\citenamefont {Lo}\ \emph {et~al.}(2005)\citenamefont {Lo},
		\citenamefont {Chau},\ and\ \citenamefont {Ardehali}}]{2005Lo}%
	\BibitemOpen
	\bibfield  {author} {\bibinfo {author} {\bibfnamefont {H.-K.}\ \bibnamefont
			{Lo}}, \bibinfo {author} {\bibfnamefont {H.}~\bibnamefont {Chau}},\ and\
		\bibinfo {author} {\bibfnamefont {M.}~\bibnamefont {Ardehali}},\ }\bibfield
	{title} {\bibinfo {title} {Efficient quantum key distribution scheme and a
			proof of its unconditional security},\ }\href
	{https://doi.org/10.1007/s00145-004-0142-y} {\bibfield  {journal} {\bibinfo
			{journal} {J. Cryptol.}\ }\textbf {\bibinfo {volume} {18}},\ \bibinfo {pages}
		{133} (\bibinfo {year} {2005})}\BibitemShut {NoStop}%
	\bibitem [{\citenamefont {Lo}\ and\ \citenamefont
		{Preskill}(2007)}]{lo2006security}%
	\BibitemOpen
	\bibfield  {author} {\bibinfo {author} {\bibfnamefont {H.-K.}\ \bibnamefont
			{Lo}}\ and\ \bibinfo {author} {\bibfnamefont {J.}~\bibnamefont {Preskill}},\
	}\bibfield  {title} {\bibinfo {title} {Security of quantum key distribution
			using weak coherent states with nonrandom phases},\ }\href
	{https://dl.acm.org/doi/10.5555/2011832.2011834} {\bibfield  {journal}
		{\bibinfo  {journal} {Quantum Inf Comput}\ }\textbf {\bibinfo {volume} {7}},\
		\bibinfo {pages} {431} (\bibinfo {year} {2007})}\BibitemShut {NoStop}%
	\bibitem [{\citenamefont {Tamaki}\ \emph {et~al.}(2018)\citenamefont {Tamaki},
		\citenamefont {Lo}, \citenamefont {Wang},\ and\ \citenamefont
		{Lucamarini}}]{tamaki2018information}%
	\BibitemOpen
	\bibfield  {author} {\bibinfo {author} {\bibfnamefont {K.}~\bibnamefont
			{Tamaki}}, \bibinfo {author} {\bibfnamefont {H.-K.}\ \bibnamefont {Lo}},
		\bibinfo {author} {\bibfnamefont {W.}~\bibnamefont {Wang}},\ and\ \bibinfo
		{author} {\bibfnamefont {M.}~\bibnamefont {Lucamarini}},\ }\bibfield  {title}
	{\bibinfo {title} {Information theoretic security of quantum key distribution
			overcoming the repeaterless secret key capacity bound},\ }\href@noop {}
	{\bibfield  {journal} {\bibinfo  {journal} {arXiv preprint arXiv:1805.05511}\
		} (\bibinfo {year} {2018})}\BibitemShut {NoStop}%
	\bibitem [{\citenamefont {Hwang}(2003)}]{PhysRevLett.91.057901}%
	\BibitemOpen
	\bibfield  {author} {\bibinfo {author} {\bibfnamefont {W.-Y.}\ \bibnamefont
			{Hwang}},\ }\bibfield  {title} {\bibinfo {title} {Quantum key distribution
			with high loss: Toward global secure communication},\ }\href
	{https://doi.org/10.1103/PhysRevLett.91.057901} {\bibfield  {journal}
		{\bibinfo  {journal} {Phys. Rev. Lett.}\ }\textbf {\bibinfo {volume} {91}},\
		\bibinfo {pages} {057901} (\bibinfo {year} {2003})}\BibitemShut {NoStop}%
	\bibitem [{\citenamefont {Beijersbergen}\ \emph {et~al.}(1993)\citenamefont
		{Beijersbergen}, \citenamefont {Allen}, \citenamefont {{van der Veen}},\ and\
		\citenamefont {Woerdman}}]{BEIJERSBERGEN1993123}%
	\BibitemOpen
	\bibfield  {author} {\bibinfo {author} {\bibfnamefont {M.}~\bibnamefont
			{Beijersbergen}}, \bibinfo {author} {\bibfnamefont {L.}~\bibnamefont
			{Allen}}, \bibinfo {author} {\bibfnamefont {H.}~\bibnamefont {{van der
					Veen}}},\ and\ \bibinfo {author} {\bibfnamefont {J.}~\bibnamefont
			{Woerdman}},\ }\bibfield  {title} {\bibinfo {title} {Astigmatic laser mode
			converters and transfer of orbital angular momentum},\ }\href
	{https://doi.org/https://doi.org/10.1016/0030-4018(93)90535-D} {\bibfield
		{journal} {\bibinfo  {journal} {Opt. Commun.}\ }\textbf {\bibinfo {volume}
			{96}},\ \bibinfo {pages} {123} (\bibinfo {year} {1993})}\BibitemShut
	{NoStop}%
	\bibitem [{\citenamefont {Jia}\ \emph {et~al.}(2018)\citenamefont {Jia},
		\citenamefont {Li}, \citenamefont {Zhang}, \citenamefont {Chen},
		\citenamefont {Wang}, \citenamefont {Gao}, \citenamefont {Li},\ and\
		\citenamefont {Zhang}}]{Jia:18}%
	\BibitemOpen
	\bibfield  {author} {\bibinfo {author} {\bibfnamefont {J.}~\bibnamefont
			{Jia}}, \bibinfo {author} {\bibfnamefont {Q.}~\bibnamefont {Li}}, \bibinfo
		{author} {\bibfnamefont {K.}~\bibnamefont {Zhang}}, \bibinfo {author}
		{\bibfnamefont {D.}~\bibnamefont {Chen}}, \bibinfo {author} {\bibfnamefont
			{C.}~\bibnamefont {Wang}}, \bibinfo {author} {\bibfnamefont {H.}~\bibnamefont
			{Gao}}, \bibinfo {author} {\bibfnamefont {F.}~\bibnamefont {Li}},\ and\
		\bibinfo {author} {\bibfnamefont {P.}~\bibnamefont {Zhang}},\ }\bibfield
	{title} {\bibinfo {title} {Integrated design of pi/2 converter and its
			experimental performance},\ }\href {https://doi.org/10.1364/AO.57.006076}
	{\bibfield  {journal} {\bibinfo  {journal} {Appl. Opt.}\ }\textbf {\bibinfo
			{volume} {57}},\ \bibinfo {pages} {6076} (\bibinfo {year}
		{2018})}\BibitemShut {NoStop}%
	\bibitem [{\citenamefont {Br{\'{a}}dler}\ \emph {et~al.}(2016)\citenamefont
		{Br{\'{a}}dler}, \citenamefont {Mirhosseini}, \citenamefont {Fickler},
		\citenamefont {Broadbent},\ and\ \citenamefont {Boyd}}]{Br_dler_2016}%
	\BibitemOpen
	\bibfield  {author} {\bibinfo {author} {\bibfnamefont {K.}~\bibnamefont
			{Br{\'{a}}dler}}, \bibinfo {author} {\bibfnamefont {M.}~\bibnamefont
			{Mirhosseini}}, \bibinfo {author} {\bibfnamefont {R.}~\bibnamefont
			{Fickler}}, \bibinfo {author} {\bibfnamefont {A.}~\bibnamefont {Broadbent}},\
		and\ \bibinfo {author} {\bibfnamefont {R.}~\bibnamefont {Boyd}},\ }\bibfield
	{title} {\bibinfo {title} {Finite-key security analysis for multilevel
			quantum key distribution},\ }\href
	{https://doi.org/10.1088/1367-2630/18/7/073030} {\bibfield  {journal}
		{\bibinfo  {journal} {New J. Phys.}\ }\textbf {\bibinfo {volume} {18}},\
		\bibinfo {pages} {073030} (\bibinfo {year} {2016})}\BibitemShut {NoStop}%
	\bibitem [{\citenamefont {Wang}\ \emph {et~al.}(2015)\citenamefont {Wang},
		\citenamefont {Sun}, \citenamefont {Ma}, \citenamefont {Tang},\ and\
		\citenamefont {Liang}}]{PhysRevA.92.042319}%
	\BibitemOpen
	\bibfield  {author} {\bibinfo {author} {\bibfnamefont {C.}~\bibnamefont
			{Wang}}, \bibinfo {author} {\bibfnamefont {S.-H.}\ \bibnamefont {Sun}},
		\bibinfo {author} {\bibfnamefont {X.-C.}\ \bibnamefont {Ma}}, \bibinfo
		{author} {\bibfnamefont {G.-Z.}\ \bibnamefont {Tang}},\ and\ \bibinfo
		{author} {\bibfnamefont {L.-M.}\ \bibnamefont {Liang}},\ }\bibfield  {title}
	{\bibinfo {title} {Reference-frame-independent quantum key distribution with
			source flaws},\ }\href {https://doi.org/10.1103/PhysRevA.92.042319}
	{\bibfield  {journal} {\bibinfo  {journal} {Phys. Rev. A}\ }\textbf {\bibinfo
			{volume} {92}},\ \bibinfo {pages} {042319} (\bibinfo {year}
		{2015})}\BibitemShut {NoStop}%
	\bibitem [{\citenamefont {Sheridan}\ \emph {et~al.}(2010)\citenamefont
		{Sheridan}, \citenamefont {Le},\ and\ \citenamefont
		{Scarani}}]{Sheridan_2010}%
	\BibitemOpen
	\bibfield  {author} {\bibinfo {author} {\bibfnamefont {L.}~\bibnamefont
			{Sheridan}}, \bibinfo {author} {\bibfnamefont {T.~P.}\ \bibnamefont {Le}},\
		and\ \bibinfo {author} {\bibfnamefont {V.}~\bibnamefont {Scarani}},\
	}\bibfield  {title} {\bibinfo {title} {Finite-key security against coherent
			attacks in quantum key distribution},\ }\href
	{https://doi.org/10.1088/1367-2630/12/12/123019} {\bibfield  {journal}
		{\bibinfo  {journal} {New J. Phys.}\ }\textbf {\bibinfo {volume} {12}},\
		\bibinfo {pages} {123019} (\bibinfo {year} {2010})}\BibitemShut {NoStop}%
	\bibitem [{\citenamefont {Wang}\ \emph {et~al.}(2016)\citenamefont {Wang},
		\citenamefont {Zhang}, \citenamefont {Wang},\ and\ \citenamefont
		{Li}}]{PhysRevA.94.062330}%
	\BibitemOpen
	\bibfield  {author} {\bibinfo {author} {\bibfnamefont {F.}~\bibnamefont
			{Wang}}, \bibinfo {author} {\bibfnamefont {P.}~\bibnamefont {Zhang}},
		\bibinfo {author} {\bibfnamefont {X.}~\bibnamefont {Wang}},\ and\ \bibinfo
		{author} {\bibfnamefont {F.}~\bibnamefont {Li}},\ }\bibfield  {title}
	{\bibinfo {title} {Valid conditions of the reference-frame-independent
			quantum key distribution},\ }\href
	{https://doi.org/10.1103/PhysRevA.94.062330} {\bibfield  {journal} {\bibinfo
			{journal} {Phys. Rev. A}\ }\textbf {\bibinfo {volume} {94}},\ \bibinfo
		{pages} {062330} (\bibinfo {year} {2016})}\BibitemShut {NoStop}%
	\bibitem [{\citenamefont {Wang}\ \emph
		{et~al.}(2021{\natexlab{a}})\citenamefont {Wang}, \citenamefont {Wu},
		\citenamefont {Dong}, \citenamefont {Zhu}, \citenamefont {Zhu},\ and\
		\citenamefont {Zhao}}]{Wang:21}%
	\BibitemOpen
	\bibfield  {author} {\bibinfo {author} {\bibfnamefont {X.}~\bibnamefont
			{Wang}}, \bibinfo {author} {\bibfnamefont {T.}~\bibnamefont {Wu}}, \bibinfo
		{author} {\bibfnamefont {C.}~\bibnamefont {Dong}}, \bibinfo {author}
		{\bibfnamefont {H.}~\bibnamefont {Zhu}}, \bibinfo {author} {\bibfnamefont
			{Z.}~\bibnamefont {Zhu}},\ and\ \bibinfo {author} {\bibfnamefont
			{S.}~\bibnamefont {Zhao}},\ }\bibfield  {title} {\bibinfo {title}
		{Integrating deep learning to achieve phase compensation for free-space
			orbital-angular-momentum-encoded quantum key distribution under atmospheric
			turbulence},\ }\href {https://doi.org/10.1364/PRJ.409645} {\bibfield
		{journal} {\bibinfo  {journal} {Photon. Res.}\ }\textbf {\bibinfo {volume}
			{9}},\ \bibinfo {pages} {B9} (\bibinfo {year}
		{2021}{\natexlab{a}})}\BibitemShut {NoStop}%
	\bibitem [{\citenamefont {Lim}\ \emph {et~al.}(2021)\citenamefont {Lim},
		\citenamefont {Xu}, \citenamefont {Pan},\ and\ \citenamefont
		{Ekert}}]{PhysRevLett.126.100501}%
	\BibitemOpen
	\bibfield  {author} {\bibinfo {author} {\bibfnamefont {C.~C.-W.}\
			\bibnamefont {Lim}}, \bibinfo {author} {\bibfnamefont {F.}~\bibnamefont
			{Xu}}, \bibinfo {author} {\bibfnamefont {J.-W.}\ \bibnamefont {Pan}},\ and\
		\bibinfo {author} {\bibfnamefont {A.}~\bibnamefont {Ekert}},\ }\bibfield
	{title} {\bibinfo {title} {Security analysis of quantum key distribution with
			small block length and its application to quantum space communications},\
	}\href {https://doi.org/10.1103/PhysRevLett.126.100501} {\bibfield  {journal}
		{\bibinfo  {journal} {Phys. Rev. Lett.}\ }\textbf {\bibinfo {volume} {126}},\
		\bibinfo {pages} {100501} (\bibinfo {year} {2021})}\BibitemShut {NoStop}%
	\bibitem [{\citenamefont {Wang}\ \emph
		{et~al.}(2021{\natexlab{b}})\citenamefont {Wang}, \citenamefont {Yin},
		\citenamefont {Liu}, \citenamefont {Wang}, \citenamefont {Chen},
		\citenamefont {Guo},\ and\ \citenamefont {Han}}]{PhysRevResearch.3.023019}%
	\BibitemOpen
	\bibfield  {author} {\bibinfo {author} {\bibfnamefont {R.}~\bibnamefont
			{Wang}}, \bibinfo {author} {\bibfnamefont {Z.-Q.}\ \bibnamefont {Yin}},
		\bibinfo {author} {\bibfnamefont {H.}~\bibnamefont {Liu}}, \bibinfo {author}
		{\bibfnamefont {S.}~\bibnamefont {Wang}}, \bibinfo {author} {\bibfnamefont
			{W.}~\bibnamefont {Chen}}, \bibinfo {author} {\bibfnamefont {G.-C.}\
			\bibnamefont {Guo}},\ and\ \bibinfo {author} {\bibfnamefont {Z.-F.}\
			\bibnamefont {Han}},\ }\bibfield  {title} {\bibinfo {title} {Tight finite-key
			analysis for generalized high-dimensional quantum key distribution},\ }\href
	{https://doi.org/10.1103/PhysRevResearch.3.023019} {\bibfield  {journal}
		{\bibinfo  {journal} {Phys. Rev. Research}\ }\textbf {\bibinfo {volume}
			{3}},\ \bibinfo {pages} {023019} (\bibinfo {year}
		{2021}{\natexlab{b}})}\BibitemShut {NoStop}%
	\bibitem [{\citenamefont {Murta}\ \emph {et~al.}(2020)\citenamefont {Murta},
		\citenamefont {Grasselli}, \citenamefont {Kampermann},\ and\ \citenamefont
		{Bruß}}]{murta2020quantum}%
	\BibitemOpen
	\bibfield  {author} {\bibinfo {author} {\bibfnamefont {G.}~\bibnamefont
			{Murta}}, \bibinfo {author} {\bibfnamefont {F.}~\bibnamefont {Grasselli}},
		\bibinfo {author} {\bibfnamefont {H.}~\bibnamefont {Kampermann}},\ and\
		\bibinfo {author} {\bibfnamefont {D.}~\bibnamefont {Bruß}},\ }\bibfield
	{title} {\bibinfo {title} {Quantum conference key agreement: A review},\
	}\href {https://doi.org/https://doi.org/10.1002/qute.202000025} {\bibfield
		{journal} {\bibinfo  {journal} {Adv. Quantum Technol.}\ }\textbf {\bibinfo
			{volume} {3}},\ \bibinfo {pages} {2000025} (\bibinfo {year}
		{2020})}\BibitemShut {NoStop}%
	\bibitem [{\citenamefont {Islam}\ \emph {et~al.}(2018)\citenamefont {Islam},
		\citenamefont {Lim}, \citenamefont {Cahall}, \citenamefont {Kim},\ and\
		\citenamefont {Gauthier}}]{PhysRevA.97.042347}%
	\BibitemOpen
	\bibfield  {author} {\bibinfo {author} {\bibfnamefont {N.~T.}\ \bibnamefont
			{Islam}}, \bibinfo {author} {\bibfnamefont {C.~C.~W.}\ \bibnamefont {Lim}},
		\bibinfo {author} {\bibfnamefont {C.}~\bibnamefont {Cahall}}, \bibinfo
		{author} {\bibfnamefont {J.}~\bibnamefont {Kim}},\ and\ \bibinfo {author}
		{\bibfnamefont {D.~J.}\ \bibnamefont {Gauthier}},\ }\bibfield  {title}
	{\bibinfo {title} {Securing quantum key distribution systems using fewer
			states},\ }\href {https://doi.org/10.1103/PhysRevA.97.042347} {\bibfield
		{journal} {\bibinfo  {journal} {Phys. Rev. A}\ }\textbf {\bibinfo {volume}
			{97}},\ \bibinfo {pages} {042347} (\bibinfo {year} {2018})}\BibitemShut
	{NoStop}%
	\bibitem [{\citenamefont {Schwonnek}\ \emph {et~al.}(2021)\citenamefont
		{Schwonnek}, \citenamefont {Goh}, \citenamefont {Primaatmaja}, \citenamefont
		{Tan}, \citenamefont {Wolf}, \citenamefont {Scarani},\ and\ \citenamefont
		{Lim}}]{Lim2021}%
	\BibitemOpen
	\bibfield  {author} {\bibinfo {author} {\bibfnamefont {R.}~\bibnamefont
			{Schwonnek}}, \bibinfo {author} {\bibfnamefont {K.~T.}\ \bibnamefont {Goh}},
		\bibinfo {author} {\bibfnamefont {I.~W.}\ \bibnamefont {Primaatmaja}},
		\bibinfo {author} {\bibfnamefont {E.~Y.-Z.}\ \bibnamefont {Tan}}, \bibinfo
		{author} {\bibfnamefont {R.}~\bibnamefont {Wolf}}, \bibinfo {author}
		{\bibfnamefont {V.}~\bibnamefont {Scarani}},\ and\ \bibinfo {author}
		{\bibfnamefont {C.~C.-W.}\ \bibnamefont {Lim}},\ }\bibfield  {title}
	{\bibinfo {title} {Device-independent quantum key distribution with random
			key basis},\ }\href {https://doi.org/10.1038/s41467-021-23147-3} {\bibfield
		{journal} {\bibinfo  {journal} {Nat. Commun.}\ }\textbf {\bibinfo {volume}
			{12}},\ \bibinfo {pages} {2880} (\bibinfo {year} {2021})}\BibitemShut
	{NoStop}%
\end{thebibliography}
\end{document}